\documentclass{article}
\usepackage{moreverb,url}
\usepackage[colorlinks,bookmarksopen,bookmarksnumbered,citecolor=red,urlcolor=red]{hyperref}

\usepackage[utf8]{inputenc}
\usepackage[english]{babel}
\usepackage{array}
\usepackage{bm} 
\usepackage{amstext,amssymb,mathtools,esint,tabularx}
\usepackage{booktabs}
\usepackage{natbib}
\usepackage{microtype} 
\usepackage{color, graphicx}

\usepackage{a4wide}
\usepackage{fixltx2e} 
\usepackage{algorithm}
\usepackage{algpseudocode}
 
\usepackage{tikz,pgfplots,pgfplotstable} 
\pgfplotsset{compat=newest}
\usetikzlibrary{calc,3d,arrows,patterns,decorations}
\tikzset{>=stealth'}
\usetikzlibrary{positioning} 
\usepackage[abs]{overpic}
\usepackage[colorinlistoftodos,prependcaption,textsize=small]{todonotes}

 \graphicspath{{./figure/}}

\newcommand{\ull}{\underline}
\newcommand{\ol}{\overline}
\def \super {\mbox{\footnotesize s}}

\def\bu{\ull{{u}}}
\def\bff{\ull{{f}}}

\def\GRB{{\bf GRB}}
\def\LRB{{\bf LRB}}

\newcommand{\dx}{\, \mathrm{d}x}

\DeclareMathOperator{\diag}{diag}

\begin{document}


\title{Adaptive control in rollforward recovery for extreme scale multigrid} 

\author{Markus Huber\footnote{Technische Universit\"at M\"unchen, Germany} and Ulrich R\"ude
\footnote{Friedrich-Alexander Universit\"at N\"urnberg-Erlangen}\footnote{CERFACS, Parallel Algorithms Project, Toulouse, France}, Germany and Barbara Wohlmuth$^{\ast}$}





\maketitle
\begin{abstract}
With the increasing number of 
compute components, 
failures in future exa-scale computer systems are expected to become more frequent.
This motivates the study of novel resilience techniques.
Here, we extend 
a recently proposed algorithm-based recovery method 
for 
multigrid iterations by introducing an adaptive control. 
After a fault, the healthy part of the system continues the iterative solution process,
while the solution in the faulty domain is re-constructed by an asynchronous on-line 
recovery.
The computations in both the faulty and healthy subdomains 
must be coordinated in a sensitive way, in particular, 
both under and over-solving must be avoided.
Both of these waste computational resources and will therefore increase the overall time-to-solution.
To control the local recovery and guarantee an optimal re-coupling, we introduce a 
stopping criterion based on a mathematical error estimator.
It involves hierarchical weighted sums of residuals within the context  of uniformly refined meshes and is well-suited in the context of parallel high-performance computing. 
The re-coupling process is steered by local contributions of the error estimator. We propose and compare two criteria which differ in their weights.
Failure scenarios when solving up to $6.9\cdot10^{11}$ unknowns on more than 245\,766 parallel processes will be reported on a state-of-the-art peta-scale supercomputer
demonstrating the robustness of the method.
\end{abstract}




\section{Introduction}
The driving force behind the research interest in achieving larger computing systems with more than $10^{18}$ floating point operations (FLOPS) per second (exa-scale) originates in the potentials to expand
science and engineering by simulation. 
For the upcoming exa-scale computing era, expected in the next decade, \cite{cappello2009,cappello2014, dongarra2011_1}, the  enormous increase in compute power also constitutes new grand challenges for these architectures.
One of the key challenges composes in building reliable systems with massively increasing components and process shrinking. This increase automatically reduces the mean time between failures (MBFT) such that using today's checkpoint/restart techniques become difficult, since the time for checkpointing and restart may exceed the expected MBFT of computing entity losses (hard faults) like cores, nodes or racks; see \cite{ropars2013,oldfield2007,cappello2009_2,cappello2009,cappello2014,shahzad2013evaluation} and references therein. In addition, when using checkpointing techniques, the additional time for each checkpoint delays the termination of the program and automatically yields a higher power consumption. Moreover, the memory necessary for checkpointing needs to be provided \cite{kohl2017scalable}
and must be kept in a reliable state, which may 
further increase energy consumption. Besides hard failures, which cause a physical loss of a computing entity, an increase of failures which are not immediately noticeable, is expected. These failures occur in "bit-flips" corrupting, e.g., floating points, and are called soft errors.
In high-end hardware 
a significant amount of power is, for example,\ consumed by error correction code (ECC) methods correcting these soft errors. ECC techniques provided by the hardware vendors significantly increase the power load of HPC-centers. Removing such power-hungry hardware design may be necessary to operate exa-scale systems and would additional make each component increasingly unreliable; see \cite{lammers2010,winstead2012,cappello2009_2} and references therein.  
Therefore, to guarantee a reliable system, high energy costs are necessary, 
%
which creates another 
grand challenge for exa-scale computing.

To minimize the overhead of reliable computation and make exa-scale computation possible by reducing the energy costs, other levels of the programming model have to maintain suitable layers of fault tolerance. Algorithm-based fault tolerance (ABFT) uses the inherent characteristics of the application  to obtain resilient results and is, therefore, very resource and energy-saving. However, it is the user's or programmer's task to define regions in the code which have to be protected or to define data that is necessary for continuing computations in case of a failure, and to implement recovery techniques. ABFT was introduced by \cite{huang1984} by using checksums to control computations and was further analyzed  in \cite{Luk1988}. Since then, strong efforts have been made in various directions to provide and incorporate reliability at the algorithmic level \cite{anfinson1988,boley1992,bosilca2009,bosilca2015}.  

One widely used parallelization model for state-of-the art high performance software uses the message passing interface (MPI). Herein, the loss of a core or compute node results in an immediate termination of the application. The direction of providing an abstraction of fault-free run-time systems to the application by, e.g., MPICH-V \cite{bouteiller2006}, redMPI \cite{fiala2012}, might be not sustainable for extreme scale \cite{heroux2013}. More promising are techniques in extending the MPI-standard in a fault-tolerant version like the user-level failure migration (ULFM) \cite{bland2012,bland2013} or the FENIX project \cite{gamell2014}, in which the ULFM MPI-extension is used for on-line recovery applications.  Through these developments, the usability of application-aware fault tolerant techniques is emphasized.

\subsection{Algorithm-based fault tolerant techniques for linear solvers}
In scientific and engineering applications, partial differential equations (PDEs) play a central role in simulations. The methods for solving linear systems arising from PDEs by discretization are of special interest for employing ABFT, since they constitute a very memory- and time-consuming part of the overall computation. 

Direct solver techniques are considered in \cite{davies2013,du2012} by using checksums to protect against soft errors. These methods are not applicable to standard iterative methods and were extended to SOR or Krylov-subspace methods in \cite{bridges2012,langou2008}. Task-based recovery of a domain decomposition solver is considered in \cite{Mycek2017a} and a priori bounds are presented for this recovery in \cite{Mycek2017}. Especially for hard errors, local recovery methods which identify the lost part of the solver as local subproblem and re-compute or re-construct a valid status to continue computations, show a certain attraction \cite{langou2008,agullo13,goeddeke2015,huber2016,gamell2017,stals2017}. 

In the following, we will use multigrid methods \cite{hackbusch1985,brandt2011},  which solve the linear system on a hierarchy on meshes, as our solver tool. These methods may have optimal complexity with respect to the number of degrees of freedom and they can be implemented very efficiently for high-performance simulations \cite{gmeiner-huber-john-ruede-wohlmuth_2015, notay2015,sundar2012,baker2016}. Therefore, ABFT techniques for soft and hard errors are of interest with these methods. The influence of soft faults on algebraic multigrid is studied in \cite{CasasSupinskiBronevetskySchulz2012}. In \cite{altenbernd2017}, soft errors are detected and corrected by checksums within multigrid schemes. 
The mathematical theory of the influence of the soft faults on multigrid schemes is considered in \cite{ainsworth2016_1,ainsworth2016_2}. Hard errors in the context of multigrid are investigated in \cite{goeddeke2015}, where the multigrid hierarchy is used to save compressed checkpoints and to restart from them. In a complete adaptive setting, the mesh hierarchy information for the multigrid solver is lost in case of core or node losses. A re-construction procedure before continuing the computation is considered in \cite{stals2017}.  In \cite{huber2016}, we introduced an ABFT technique for resilient multigrid methods for hard failures. More specifically, we have focused on a local recovery method within multigrid V-cycles and combined it with domain decomposition ideas to develop a global recovery method. We have been able to show that the overhead caused by faults in terms of run-time can be completely compensated when accelerating the local recovery by additional compute power, i.e., by using a {\em superman} and by selecting a suitable number of iterations in the recovery.


\subsection{Contribution}
In this article, we extend the global recovery idea proposed in \cite{huber2016}, in which we fixed {\it a priori} the number of iterations in the recovery. The quality of the recovery depends sensitively on this number. Thus, there is a need for an automatic selection based on a rigorous criterion. Here, we use an adaptive {\it a posteriori} control mechanism for the algebraic error, which steers the synchronization of the asynchronous algorithm and re-couples faulty and healthy processors. This extension is of interest, since remaining in the recovery mode for too long increases the run-time without improving the approximation towards the solution. On the other hand, interrupting the recovery process too early leads to a pollution of the approximation and additional multigrid iteration to converge to a sufficient approximation. Therefore, the total run-time increases here too. A sophisticated re-coupling strategy is of importance to allow a run-time minimal recovery. Our automatic re-coupling is based on controlling the approximation quality in the faulty domain by a threshold.
that is based on the algrebaic error before the fault has occured.
We apply a mathematically motivated error estimator for both components of the criterion to guarantee an efficient re-coupling.  Many error estimator concepts exist in the mathematical literature, through most of them are too costly for high-performance computing and proposed for the discretization error or the quantity of interest; see \cite{ainsworth2000}.
Thus, we will consider the hierarchical weighted (HW) residual estimator (see \cite{ruede1993_1,ruede1994}) which employs typical multigrid components to estimate the algebraic error and can be implemented with minimal overhead within multigrid schemes. In addition, the estimator very accurately represents the error locally. 

All these features of the estimator make it useful to employ it for our recovery. 
Depending on the scenario described, it may be beneficial to use different stopping strategies. For the definition of the re-coupling bounds, we distinguish between the following two strategies. The  {\it global re-coupling bound} enforces to solve the local subproblem to the size of the algebraic error before the fault. The  {\it local re-coupling bound} requires that the approximation quality after the recovery fulfills a local condition 
which can be interpretaded as weighting the global error before the fault by the volume ratio associated with a single process.

Both criteria can be evaluated in case of a failure by a single global communication and without additional computations such that the overhead is kept to a minimum.
In the recovery, an error indicator motivated by the HW estimator can be applied to obtain a quantitative error representation in the faulty domain. When this indicator satisfies the stopping criterion, the recovery is terminated and faulty and healthy subproblems are re-coupled. It has been verified through numerical experiments that the indicator guarantees an optimal recovery method without any run-time overhead.
Since the indicator is used in the faulty domain, it can also profit from the superman acceleration. We test the adaptive {\it a posteriori} recovery strategy on a state-of-the art supercomputer for the stopping criterion with two re-coupling bounds within different fault scenarios. We also show that our method is flexible with respect to MBFT 
for the case that two failures occur during the multigrid iteration.


The paper is organized as follows: In Section \ref{sec:setup_model}, we introduce the model problem, the discretization, the multigrid solver and the data structures. Then, we define in Section \ref{sec:faulty} the settings for the faulty computing environment. In Section \ref{sec:exemplaryfault}, we introduce an exemplaric fault scenario for which we numerical study the impact of core losses on the multgrid convergence and the recovery algorithm. In Section \ref{sec:global}, we present the principles of the global recovery strategy and extend this in Section \ref{sec:adaptiveStrat} to an adaptively controlled recovery strategy. In Section \ref{sec:estimator}, we introduce an explicit choice for indicating the error in the faulty domain and the different stopping criteria for re-coupling after the local recovery. Both are based on the hierarchical weighted algebraic error estimator. We present several numerical experiments. Finally, we validate the recovery algorithm by extensive numerical experiments on a state-of-the-art peta-scale supercomputer in Section~\ref{sec:numerics}. We consider the multigrid convergence in a single failure, when adaptively controlled recovery algorithm is applied in Section~\ref{sec:numericsexperimentsSingle}. We compare the run-time performance of the involved estimators in scaling experiments in Section~\ref{sec:perform2}. The asymptotic behavior of the re-coupling criterion with global and local bounds are then study in terms of run-time overhead in Section~\ref{sec:performance_recoupling2}. The local error contribution of each process befor and after the failure and after the recovery is considered in Section~\ref{sec:local_dist}. In Section~\ref{sec:numericsexperiments}, we verify that our recovery algorithm is robust and relexible useable with respect to MBFT variation.

\section{Model problem, multigrid solver and data-structure\label{sec:setup_model}}
%
%
In this section, we introduce the model problem, the discretization by a standard low order finite element method and the multigrid solver. The implementation of the multigrid method is carried out within the high-performance {\it hierarchical hybrid grids} (HHG) framework \cite{bergen2006,gmeiner-huber-john-ruede-wohlmuth_2015}. The underlying data structure of the framework is common to many parallel multigrid implementations, \cite{huelsemann2005, falgout2000} and constitutes no restriction for the proposed algorithm-based fault tolerance. Our techniques can be realized within other frameworks such as described in  \cite{sundar2012,baker2016,petsc2016}.
\subsection{Model problem\label{sec:model}}
We consider a bounded polyhedral domain $\Omega \subset \mathbb{R}^3$ and denote by $u \in V\coloneqq H^1_{0}(\Omega)$ the solution of the Poisson equation
\begin{equation}\label{eq:model}
\begin{alignedat}{3}
-\Delta u &= f &\qquad& \text{in } && \Omega,\\
u &= 0 && \text{on } &&\partial\Omega,
\end{alignedat}
\end{equation}
with 
$f \in L^2(\Omega)$, where $L^2(\Omega)$ is the space of all square integrable functions on $\Omega$,
and $H^1_0(\Omega)$ is the standard Sobolev space  with zero trace on the boundary.
The homogeneous boundary condition in \eqref{eq:model} will simplify the notation; a generalization to inhomogeneous and more general boundary conditions is easily possible, see Section \ref{sec:numerics}. The weak formulation of \eqref{eq:model} then reads: find $u \in V$ such that it satisfies
\begin{equation}\label{eq:modelweak}
a(u,v) = f(v)\qquad \forall v\in V,\\
\end{equation}
where $a(\cdot,\cdot)$ is the bilinear form 
\begin{equation}\label{def:bilinear}
a\colon V \times V \to \mathbb{R},\quad a(u,v) \coloneqq \int_{\Omega} \nabla u \cdot \nabla v \dx,
\end{equation} 
and  
$f(\cdot)$
a linear functional
\begin{equation}
f\colon V \to \mathbb{R},\quad f(v) \coloneqq \int_{\Omega} f \, v \dx.
\end{equation}
%
%
We discretize \eqref{eq:modelweak} using standard conforming linear finite elements on a hierarchy of uniformly refined meshes $\mathcal{T}:=\{\mathcal{T}_{\ell}, ~\ell=0,...,L$\}, where $\ell$ defines the level of refinement, and $\mathcal{T}_L$ is the finest mesh. The corresponding finite dimensional approximation spaces are denoted by $V_0\subset V_1 \subset ....\subset V_L\subset V$. Using the standard nodal basis functions defined by $\psi_{\ell,j}$ for $j=1,\dots,n_\ell$  on level $\ell$, the isomorphism $v \leftrightarrow \ull v$ maps $v \in V_\ell$ to the coefficient vector $\ull v \in \mathbb{R}^{n_\ell}$, where $n_\ell = \text{dim}(V_\ell)$. The systems of linear algebraic equations corresponding to the finite element discretization on level $\ell$ of \eqref{eq:modelweak} read as
\begin{equation}\label{eq:modelLS}
A_\ell \bu_\ell = \bff_\ell, \quad \ell=0,\dots,L,
\end{equation}
with entries 
\begin{equation}
(A_\ell)_{ij} = a(\psi_{\ell,j},\psi_{\ell,i}), \quad (\bff_\ell)_j = f(\psi_{\ell,j})
\end{equation}
for $i,j =1,\dots, n_\ell$, and ${\bu}_\ell$ is the coefficient vector of the finite element solution $u_\ell$ on level $\ell$.

Note, while  in the following  tetrahedral meshes are used, all our techniques generalize to hexahedral and hybrid meshes:
\subsection{Solver setup\label{sec:MGsolver}}
Multigrid is known for its mesh-independent convergence and optimal complexity in the number of unknowns \cite{hackbusch1985,brandt1994}. Therefore, parallel multigrid is of a special interest for large scale high-performance computations, \cite{gmeiner-huber-john-ruede-wohlmuth_2015,notay2015,sundar2012,rudi2015}. The mesh hierarchy $\mathcal{T}$ introduced in Section \ref{sec:model} is used to construct a geometric multigrid solver with its coarsest level for $\ell=0$ and finest level for $\ell=L$. In order to solve the algebraic equation $A_L \ull u_L =\ull f_L$ associated with the finest mesh, we apply multigrid V-cycles; see, e.g, \cite[Algorithm (2.5.4)]{hackbusch1985}. The intergrid transfer operators are realized by linear interpolation $I_\ell^{\ell+1}$ and the adjoint operator for restriction is defined by $I_{\ell+1}^\ell := (I_\ell^{\ell+1})^{\top}$ for $\ell=0,\dots,L-1$.  We choose the operators in \eqref{eq:modelLS} for level $\ell=0,\dots,L-1$ defined by direct assembly on each level $\ell$ as coarse grid operators within the V-cycle approximation scheme. We note that for this choice of transfer operators and conforming linear finite element discretization for the model problem \eqref{eq:model}, the operators defined in \eqref{eq:modelLS} for level $\ell=0,\dots,L-1$ coincide with the Galerkin approximation, i.e., 
\begin{equation}\label{def:CoarseGridOp}
A_\ell = I_L^\ell A_L I_\ell^L,
\end{equation} 
where $I_\ell^L = I_{L-1}^L I^{L-1}_{L-2}...I_{\ell}^{\ell+1}$ and $I_L^\ell = (I_\ell^L)^\top$. For smoothing, we use standard point-wise relaxation routines, e.g., (damped) Jacobi or (colored) Gauss-Seidel smoothers. The parallel implementation is based on a hybrid realization of the smoothers; for details, see \cite{gmeiner-huber-john-ruede-wohlmuth_2015}.

In the following, we will study the convergence process of an approximation $\ull u_L^k$ to the exact solution $\ull u_L$ of \eqref{eq:modelLS} on level $L$ within a sequence of V-cycle iterations $k=0,1,2,\dots$. We study here the algebraic error, i.e.,
\begin{equation}
\ull e_\ell^k \coloneqq \bu_\ell - \bu_\ell^k,
\end{equation}
and measure it in a weighted Euclidean norm
\begin{equation}\label{def:L2error}
\|\ull e_L^k\|_{0;L}
\end{equation}
 equivalent to the weighted (by $|\Omega|^{-1/2}$) $L^2$-norm (the equivalence constant only depends on the shape regularity of the triangulation $\mathcal{T}_\ell$) defined by  
\begin{equation}\label{def:discreteL2norm}
 \|\ull w_L \|_{0;L}^2 := \frac{1}{n_L} \ull w_L^\top \ull w_L, \quad \ull w_L\in\mathbb{R}^{n_L}.
\end{equation}
We call \eqref{def:discreteL2norm} therefore also discrete $L^2$-norm. In what follows, we will focus on an open subdomain $\omega\subset\Omega$, i.e., we will be working with a subset $\mathcal{I}_{\ell;\omega}$ of degree of freedoms (DOFs) contained in $\omega$ of the index set of all DOF $\{1,\dots,n_\ell \}$
\begin{equation}\label{eq:subdomainindex}
\begin{aligned}
\mathcal{I}_{\ell;\omega} \coloneqq \{& i\in\{1,\dots,n_\ell\},\\ &i \text{ is DOF on level $\ell$ located in } \omega\}.
\end{aligned}
\end{equation}
for $\ell=0,\dots,L$.
On this subset, we define the discrete weighted $L^2$-norm by
\begin{equation}\label{def:discreteL2normIndex}
 \|\ull w_\ell\|_{0;\mathcal{I}_{\ell;\omega}}^2 \coloneqq \frac{1}{n_{\mathcal{I}_{\ell;\omega}}} \sum \limits_{j\in \mathcal{I}_{\ell;\omega}} (\ull w_\ell)_j^2,
\end{equation}
where $n_{\mathcal{I}_{\ell;\omega}}$ is the dimension of the subset. Due to the weighting in \eqref{def:discreteL2normIndex}, the norm is not additive with the respect to subdomains. The level index in the subset \eqref{eq:subdomainindex} will be dropped below, when the considered level is clear or not of importance. Note, these subsets of DOF will be chosen such that they correspond to (sub-)domains of interest, i.e., faulty or healthy domain, of $\Omega$. 

Since the exact solution $\bu_L$ of \eqref{eq:modelLS} within an iterative method is not available for evaluating \eqref{def:L2error}, it is therefore commonly accepted to use the  residual of  \eqref{eq:modelLS}  on level $L$
\begin{equation}\label{def:residual}
\ull r_L^k = \ull f_L - A_L \ull u_L^k.
\end{equation}
instead to stop the approximation process. Here, the relation between algebraic error and residual
\begin{equation}\label{def:residual_algerror}
\ull r_L^k = A_L \ull e_L^k
\end{equation}
is inherently used. However, as we will consider in more detail in Section \ref{sec:HWestimator}, this does not allow to bound the algebraic error independent of the mesh size $h_L$ and only for $\ull r_L^k = 0$, it is equivalent to {$\ull e_L^k = 0$}. Therefore, we also consider a more sophisticated error estimator for the algebraic error in the following. We terminate the approximation process, when the relative criterion for the error indicator $\eta_{\Omega}^k$ after $k$ steps of the iterative method has been reduced  
%
by a specified tolerance $\text{TOL}$, i.e.,
\begin{equation}\label{def:stoppingcriterionGlobal}
\eta_{\Omega}^k < \text{TOL} \,\, \eta_{\Omega}^0.
\end{equation}
%

%

\subsection{Data structure principles\label{sec:HHGdatastructure}}

In this section, we  briefly discuss the data structures that enable efficient multigrid computations and are suitable for the following ABFT recovery strategy. Conceptually, a hybrid data structure is used which combines multigrid mesh hierarchies with tearing and interconnecting strategies from domain decompositions and conforms to a commonly used implementation of parallel multigrid \cite{huelsemann2005,falgout2000}. All our results will be carried out in the hierarchical hybrid grids (HHG) framework, see \cite{bergen2006,gmeiner2015_2}, which constitutes such a software package employing this hybrid data structure. Moreover, HHG can be seen as representative test environment for the realization of ABFT methods, since it has demonstrated the capability to scale to a very large processor number for application-orientated science \cite{gmeiner-huber-john-ruede-wohlmuth_2015, weismueller_15}.  

\begin{figure}[!htb]
\centering
\includegraphics[width=0.47\textwidth]{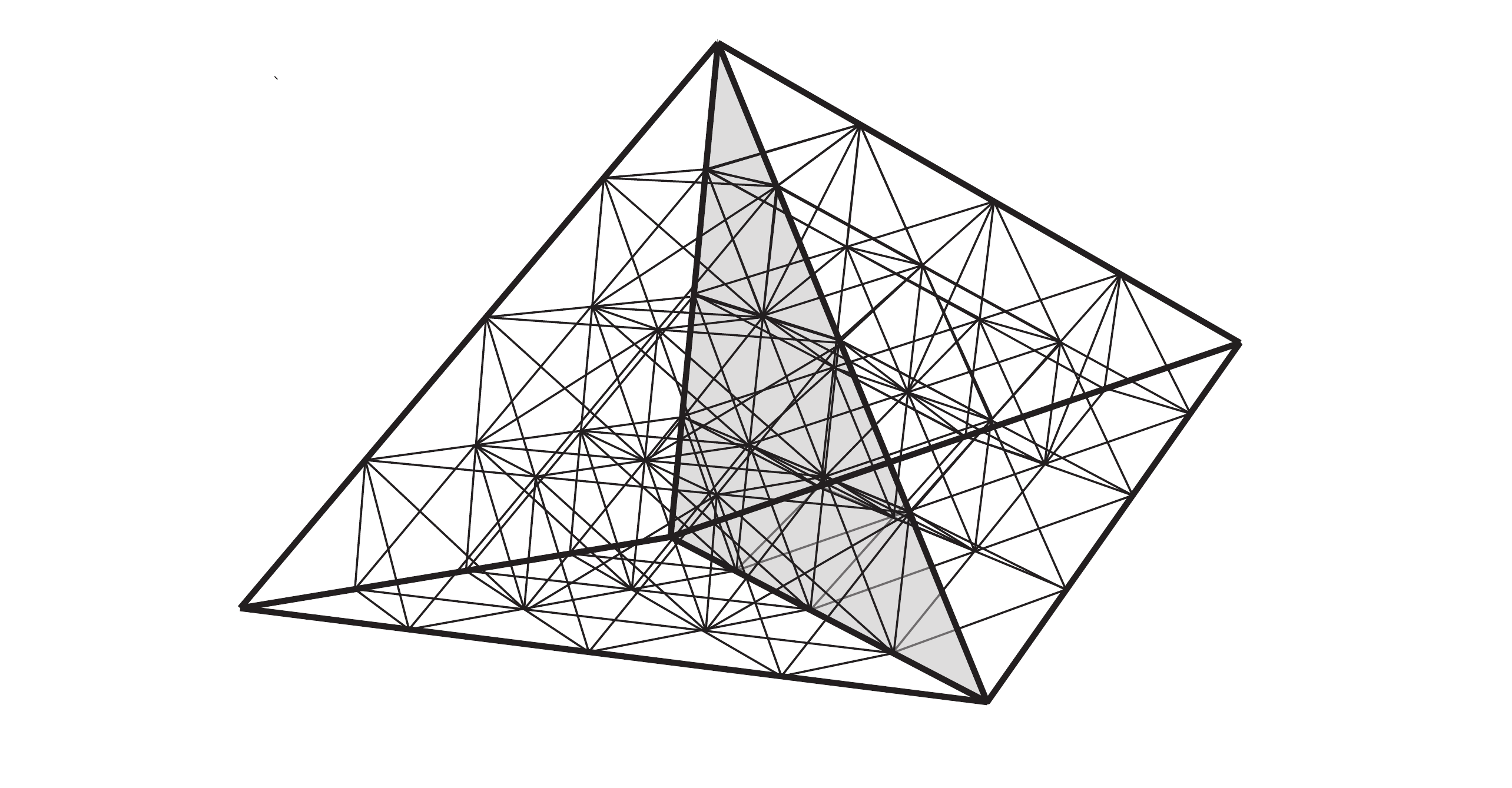}\qquad
\includegraphics[width=0.47\textwidth]{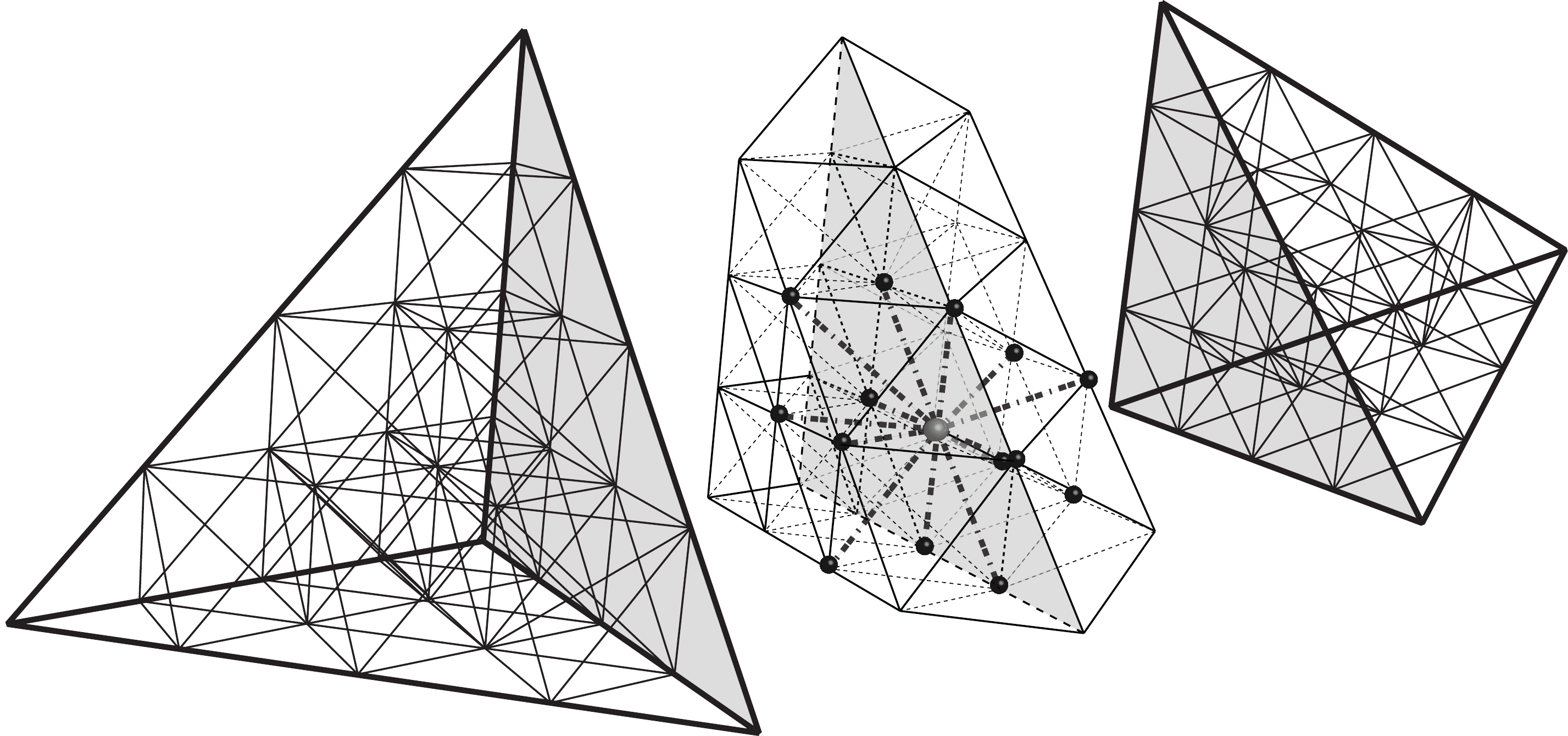}
\caption{top: two refined macro-elements; bottom: ghost layer structure of two macro-elements.\label{fig:HHGghostlayer}}
\end{figure}
An efficient  communication model is to introduce a layer of halos (ghost layers) that hold dependent redundant copies of {\em master} data on other memory units. Data dependencies can therefore be organized  across process boundaries and make parallel computations possible. The data on these ghost layers can only be read such that it needs to be updated when the master data is modified to hold consistent values.  Here, a systematic construction of a ghost layer data structure will be highlighted which is induced by the geometry of the mesh. 
A hierarchically organized data structure and communication model can be defined on top of the uniform refined meshes in the family $\mathcal{T}$. 
For each of these meshes, the nodal values of a discretization level lie on the vertices, edges, faces, and in the interior of an initial (macro-) element of the input mesh $\mathcal{T}_0$. For two tetrahedra in 3D, this is illustrated in Figure \ref{fig:HHGghostlayer} (left). We classify these nodal values according to a system of container data structures. Interior nodal values of a macro-element $T_i \in \mathcal{T}_0$ are stored in a 3D container, the nodal values lying on a input mesh face $\ol F_{i,j} = \ol T_i \cap \ol T_j$ for $T_i,T_j\in\mathcal{T}_0$ in a 2D container, the values on a input mesh edge in 1D container and the nodal values of the input mesh vertices in a 0D container. 
In a subsequent step, we introduce the ghost layers. Through the above geometric classification, each nodal value is stored in an unique container and is referred as {\em master} copy. These containers are now enriched with copies of neighboring nodal values stored as master data in another container, which are called {\em ghost} values. Thus, for an element $T_i \in \mathcal{T}_0$ all boundary nodal values become ghost values. For the face data structure, the values which lie on the boundary of the face and additional layers holding the values of the adjacent tetrahedra define the ghost layer nodal values. 
Edges and vertices are enriched similarly.
In Figure \ref{fig:HHGghostlayer} (right), the ghost layer enrichment for two macro-tetrahedra and the face lying in between is illustrated. These extra ghost layers are essential for the efficient implementation of local operations such as relaxation methods acting on the master nodes.

The parallelization on a distributed memory system is based on the distribution of the containers uniquely to processors. If the master data of an interface data structure (face, edge, vertex) has changed on a processor, it needs to be updated in the ghost layers. Depending on the memory location, the update is a local copy routine if the container belongs to the same memory unit, or message passing (MPI) communication, if the containers are located on different memory units in the network.
Here, we eventually introduce additional 
copies of the face, edge, and vertex containers so that the communication can be efficiently implemented on the processors of a distributed memory system.

This construction obviously leads to redundancy in face, edge and vertex data. However, the extra stored data is of lower order complexity and induces a negligible memory overhead in the typical cases. 
It makes it possible to implement a very efficient parallel multigrid solver and permits the numerical recovery of the data in these structures.
In case of a failure, the loss of master face, edge and vertex data can be recovered by the redundant information on other processes. Therefore, we focus in the following only on the algorithmic recovery of the macro-element unknown data.

\section{Resilience for the multigrid solver\label{sec:global}}

In this section, we introduce the model for dealing with faults within a many-core computing environment from \cite{huber2016}. 
One strategy for dealing with faults within multigrid schemes is based on tearing and interconnecting concepts from domain partitioning, and it decouples faulty and healthy parts to solve Dirichlet problems in the corresponding subproblems. 
This is the idea of the Dirichlet-Dirichlet recovery strategy considered our paper \cite[Algorithm 2]{huber2016}.
The extension of this algorithm to an adaptive controlled strategy is then considered in the forthcoming sections.
The impact of a fault on the iteration process of multigrid methods is studied by considering an exemplary fault situation within the V-cycle schemes.
This showcase will also serve in the following to study  the adaptive controlled recovery algorithm.

%
\subsection{Faulty computing environment\label{sec:faulty}}
In the following, we consider hard failures, i.e., the loss of a computing entity such as one or several cores, such as a complete compute node. We call these entities {\it faulty processor units} and the rest {\it healthy processor units}. The immediate detection and instant replacement of faulty processor units is not in the standard of the messaging passing interface (MPI), and not yet routinely supported by hardware and software. However, efforts are under way.
Currently, only extensions such as the user-level-failure-migration (ULFM) are available \cite{bland2012,bland2013}. Since an instant detection and replacement of the faulty process is essential for our method, we only simulate the fault and do not explicitly kill the processes. The data of faulty entities is lost and has to be recovered.  We distinguish between static data and dynamic data recovery. Static data in what follows like internal data-structure, stiffness matrix or right-hand side has to be recovered by standard check-pointing methods or new setup. We study the recovery of the dynamic data (unknown vector) numerically and set the nodal values of the lost entities to zero, when injecting the fault. Since we only consider settings in which the distributed memory parallelization is based on the input mesh $\mathcal{T}_0$ and the uniform refined meshes are distributed according to $\mathcal{T}_0$, we can identify crashed processes with elements of $\mathcal{T}_0$ assigned to them, see also Section \ref{sec:HHGdatastructure}. Thus, when a process crashes, the data of nodal values of the hierarchy of these tetrahedrons is lost and must be recovered. The simulation of faults in a computing environment is a wide research topic, see e.g., \cite{Hsueh1997, Benso2010} and is beyond our consideration. Therefore, we artificially inject a fault into our simulation and state that the data of specific processes are lost.
For instance in Fig.~\ref{fig:FaultTet}, one process crashes and the nodal data values of one refined macro-tetrahedron (marked red) is lost. The data of the part of the computational domain $\Omega$ which is lost due to the failure is called {\it faulty} domain $\Omega_F$. The subdomain $\Omega_I = \Omega \backslash \overline{\Omega}_F$ is called {\it healthy} or {\it intact} domain (marked blue).  Further, we define the {\it interface} $\Omega_{\Gamma}:=\overline{\Omega}_F \cap \overline{\Omega}_I$, i.e., the nodes which have dependencies in the faulty and healthy domain are located on $\Omega_\Gamma$. We decompose a vector $\ull v \in \mathbb{R}^{n_L}$ to $\ull v_{\mathcal{I}_{\Omega_F}}$, $\ull v_{\mathcal{I}_{\Omega_I}}$ and $\ull v_{\mathcal{I}_{\Omega_{\Gamma}}}$ denoted with respect to their subdomains, cf. Fig. \ref{fig:FaultTet} and \eqref{eq:subdomainindex} for the definition of the index subset of all DOF. 

\begin{figure*}[!ht]
  \include{FaultDomain}
   \caption{Domain $\Omega$
	 with a faulty subdomain $\Omega_{F}$, interface $\Omega_{\Gamma}$ and healthy subdomain $\Omega_I$.}
   \label{fig:FaultTet}
\end{figure*}

\subsection{Exemplary fault recovery environment\label{sec:exemplaryfault}}
 
We assume a failureoccurs after $k=k_F$ iterations during the approximation within multigrid V-cycles. For simplicity, we only consider failure after a full V-cycle, i.e., the fault is injected after post-smoothing and before pre-smoothing on the finest grid level.  This can be easily generalized by stopping the calculation when a failure is notified and returning immediate back to the top level of the V-cycle. Also other variants are possible, that, e.g., complete the multigrid cycle in the healthy domain and then start the recovery.

\begin{figure}[!htb]
  \include{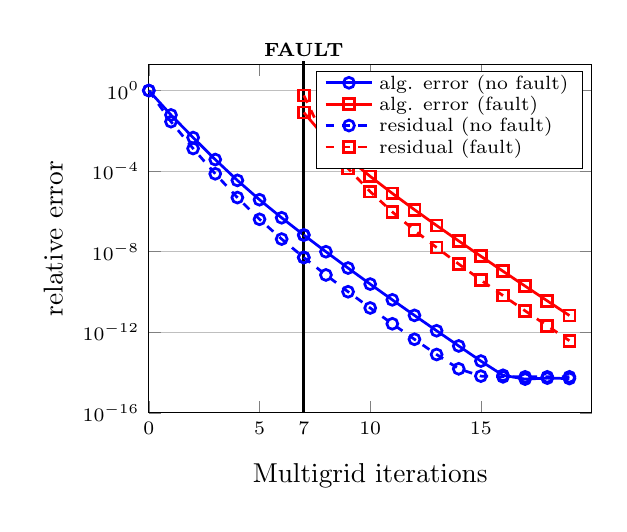}
   \caption{Iteration process of a V(3,3)-cycle with a failure after $k_F=7$ iterations; relative algebraic $L^2$-error and relative $L^2$-norm of the residual for no-fault and fault cases.}
   \label{fig:fautl_recovery}
\end{figure}

Due to the failure, the lost values of the unknown vector $\ull u_{\mathcal{I}_{\Omega_F}}^{k_F}$ in the fault domain are set to zero and, then, the multigrid scheme continues or a recovery strategy is started. In the case of a recovery, we again apply multigrid V-cycles to the overall problem after it has finished.
In Fig. \ref{fig:fautl_recovery},  we consider as model problem \eqref{eq:model} in $\Omega=(0,1)^3$ with exact solution
\begin{equation*}
u=\sin((x+\sqrt{2}y)\pi)\sin(\sqrt{3}z\pi).
\end{equation*}
 The right-hand side is 
 and Dirichlet boundary condition are prescribed according to the exact solution. 
  We apply V-cycles with three pre- and post-smoothing steps of a hybrid parallel Gauss-Seidel relaxation scheme, see Section \ref{sec:MGsolver}, denoted by V(3,3)-cycle, to solve the resulting linear system of equations \eqref{eq:modelLS}  and inject a fault after $k_F=7$ iterations. We depict both the relative  error and the relative residual with respect to the algebraic $L^2$-norm \eqref{def:discreteL2norm} for a fault-free and faulty execution. In the event of a fault, the algebraic error and the residual jump to the size of the initial algebraic error and the residual, respectively. After the failure, a pre-asymptotic convergence is observed for two iterations, but the delay counted by the number of additional needed V-cycle iterations to reach the same accuracy for the fault-free execution and after injecting a failure is 
  6-7 iterations.
%
\section{Adaptive re-coupling fault recovery strategy}

\begin{figure}[ht!]
%
\centering
	\pgfplotsset{tick label style ={font=\scriptsize}}
	\begin{tikzpicture}[scale=0.79]
		\begin{semilogyaxis}[
		width=0.6\textwidth,
		height=0.5\textwidth,
		xlabel={Multigrid iterations},
		ylabel={relative error},
		xticklabels = {0, 5, 7, 10, 15},
		xtick = {0, 5,7, 10, 15},
		ymajorgrids = true,
		xmin = 0, xmax = 20,
		ymin=1e-16,ymax=1e0,
		ytick = {1e0, 1e-4, 1e-8, 1e-12, 1e-16},
		legend style={/tikz/every odd column/.style={yshift=2pt}, yshift=0.0em, xshift=0.0em, font=\scriptsize, text height = 0.5ex, legend cell align = left},
		clip=false
		]
		
\def \sh {0}

		 \addplot[color=blue,mark=o,line width=1pt] coordinates {

(0, 1.00e+00)
(1, 4.32e-02)
(2, 2.39e-03)
(3, 1.81e-04)
(4, 1.70e-05)
(5, 1.80e-06)
(6, 2.13e-07)
(7, 2.80e-08)
(8, 4.01e-09)
(9, 6.17e-10)
(10, 9.98e-11)
(11, 1.68e-11)
(12, 2.90e-12)
(13, 5.15e-13)
(14, 9.36e-14)
(15, 1.79e-14)
(16, 7.22e-15)
(17, 8.08e-15)
(18, 8.16e-15)
(19, 7.98e-15)
             };

 \addplot[color=red,mark=o,dashed, mark options = {solid}, line width=1pt] coordinates {
(7, 8.12e-02)
(8, 1.30e-06)
(9, 1.03e-08)
(10, 1.01e-08)
(11, 1.01e-08)
(12, 1.01e-08)
(13, 1.01e-08)
(14, 9.90e-10)
(15, 1.25e-10)
(16, 1.81e-11)
(17, 2.84e-12)
(18, 4.72e-13)
(19, 8.22e-14)
              };
              

	\fill[text=FAULT, blue,fill opacity=0.25, ] (axis cs:7,1e-16) rectangle (axis cs: 7+\sh,1e2);  
	\draw[color=black,line width=1pt] (axis cs:7,1e-16) to (axis cs:7,1e0);            

	\node[rotate=0] at (axis cs:7,1e1) { \textbf{ \scriptsize FAULT}};             %
          

   \addplot+[color=red, mark=square, mark options={solid}, line width=1pt] coordinates { 
(7, 8.12e-02)
(8, 1.30e-06)
(9, 1.22e-07)
(10, 1.68e-08)
(11, 2.63e-09)
(12, 4.41e-10)
(13, 7.74e-11)
(14, 1.39e-11)
(15, 2.56e-12)
(16, 4.78e-13)
(17, 9.00e-14)
(18, 1.77e-14)
(19, 5.75e-15)
            };
%
%
	\legend{ fault-free,  over-solving, under-solving}

 	\end{semilogyaxis}
	\end{tikzpicture}
 \caption{\label{fig:underoversolving}Iteration process of a V(3,3)-cycle with a failure after $k_F=7$ iterations; relative algebraic $L^2$-error, when applying the Dirichlet-Dirichlet recovery algorithm with a suboptimal paring $(n_F,n_I)$.}
\end{figure}
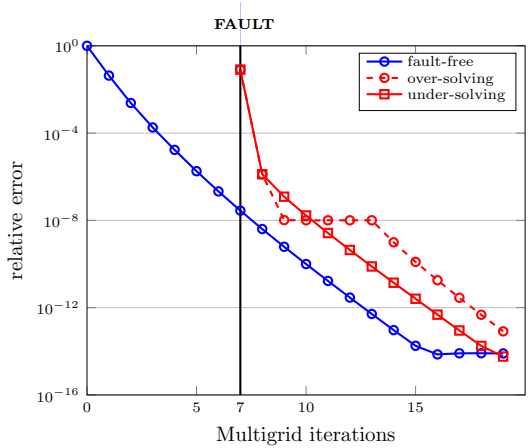

In this section, we extend the Dirichlet-Dirichlet algorithm introduced in  \cite{huber2016} by an adaptive re-coupling startegy. In case of a failure, the algorithm decouples faulty and healthy processes and solves independent Dirichlet problems on both associated subdomain by fixing the interface data. Therefore, communication between healthy and faulty subdomains is avoid and the locally increased error does not pollute the unknown values of the healthy domain. 
To compensate for the resulting error difference between faulty and healthy domain and to obtain a nearly equally distributed error after the recovery, more computations are necessary in the faulty domain. Therefore, we introduce more compute power in the faulty domain in form of a superman speed up factor $\eta_{\super} \in [1,\infty)$ to account for this imbalance. This can be realized by a further domain decomposition of the faulty domain and assigning more compute processes to the faulty domain which were handled by single process before the fault. Hence, the superman $\eta_{\super}$ is the ratio between the compute power in the faulty domain after and before the fault. 
Due to the wrong interface data, we have to re-couple both subproblems to solve the global problem. 
In order to guarantee almost equal execution times in healthy and faulty domain, i.e., $t_F \approx t_I$, and sufficiently recovery of the data in the faulty domain, the performance sensitively depends on the choice of executed iterations $n_F$ and $n_I$ in the faulty and healthy domain. It is possible to identify an optimal paring $(n_F,n_I)$ depending on the superman factor $\eta_{\super}$ such that the delay in convergence introduced by the failure can be completely compensated. A suboptimal choice of these parameters lead on the one hand an unsufficient re-covery of the data in the faulty domain, we call that {\it under-solving}. On the other hand, by remaining to long in the recovery, the global approximation obtained by subproblems suffers from the wrong interface data, we call that {\it over-solving}. Both waste computational resources and time and therefore delay the convergence. For instance in Figure \ref{fig:underoversolving}, the Dirichlet-Dirichlet algorithm is applied with suboptimal $(n_F,n_I)$ paring. The one yields over-solving  with $(n_F,n_I)= (4,1)$ and the other  under-solving with $(n_F,n_I)=(16,4)$ in the recovery and both a delay in convergence of 2-3 iterations.

By heuristics, these parameters are set {\it a priori} before applying the recovery strategy and are influenced by several problem and solver dependent factors as the coarse level solver. Therefore, an {\it a posteriori} choice of these parameters is necessary to find optimal parameter parings without extensive numerical studies determine them.

\subsection{Adaptive re-coupling algorithm\label{sec:adaptiveStrat}}
In the following, we extend Dirichlet-Dirichlet algorithm of \cite[Algorithm 2]{huber2016} by a re-coupling strategy for the faulty and healthy subproblem when the faulty problem is sufficiently approximated. Therefore, we shortly re-call some components of the algorithm. We re-write the linear system $A_L \ull u_L = \ull f_L$, see \eqref{eq:modelLS} in terms of Dirichlet problems on the subdomains and add the ghost-layer data-structures (see Section \ref{sec:HHGdatastructure}) at the interfaces to represent the necessary coupling conditions such that the stiffness matrix is given by
%
\begin{equation}\label{eq:algSystemGhost}
\small
A_L = 
 \begin{pmatrix}
  A_{II}  & A_{I\Gamma_I} & {\bf 0} & {\bf 0} & {\bf 0} \\
  {\bf 0} & \text{\bf Id} & -\text{\bf Id} & {\bf 0} & {\bf 0}\\
  A_{\Gamma I} & {\bf 0} & A_{\Gamma \Gamma} & {\bf 0} & A_{\Gamma F} \\
  {\bf 0} & {\bf 0} & -\text{\bf Id} & \text{\bf Id} & {\bf 0} \\
   {\bf 0} & {\bf 0} & {\bf 0} & A_{F \Gamma_F} & A_{F F} \\
 \end{pmatrix}
 \end{equation}
 and the unknown and the right-hand side vector by
 \begin{equation}
 \ull u_L =
 \begin{pmatrix}
  \ull u_{\mathcal{I}_{\Omega_I}}\\
  \ull u_{\mathcal{I}_{\Omega_{\Gamma_{I}}}}\\
  \ull u_{\mathcal{I}_{\Omega_{\Gamma}}}\\
  \ull u_{\mathcal{I}_{\Omega_{\Gamma_{F}}}}\\
  \ull u_{\mathcal{I}_{\Omega_F}}\\
 \end{pmatrix},
 \quad \ull f_L =
  \begin{pmatrix}
  {\ull f_{\mathcal{I}_{\Omega_I}}}\\
   {\bf 0}\\
   {\ull f_{\mathcal{I}_{\Omega_\Gamma}}}\\
   {\bf 0}\\
  \ull f_{\mathcal{I}_{\Omega_F}}\\
 \end{pmatrix} .
\end{equation}
The sub-matrices are associated with the block unknowns. In more general settings, they also depend on the basis functions and the PDE. The sub-matrices $A_{XX}$ with $X\in \{F,\Gamma, I\}$ are the blocks of the stiffness matrix $A_L$ corresponding to the parts of the vector $\ull u_L$ with dependencies just in $\Omega_X$.  The block matrices $A_{XY}$ with $X\in\{F,\Gamma,I\}$ and $Y\in \{\Gamma_I,I,F,\Gamma_F\}$, $X\neq Y$, represent the couplings in the stiffness matrix between the subdomains $\Omega_F, \Omega_I$ and $\Omega_\Gamma$. Because of the symmetry of the bilinear form \eqref{def:bilinear}, we can identify $A_{\Gamma I}$ with $A^{\top}_{I\Gamma_I}$ and $A_{\Gamma F}$ with $A^\top_{F\Gamma_F}$.
The consistency of the redundant data across process boundaries is guaranteed in the linear system \eqref{eq:algSystemGhost} by row 2 and row 4. The coupling of the interface data in healthy and faulty domains is presented by row 3. 
In case of a fault, the data corresponding to the sub-domain $\Omega_F$ is lost. However, interface data is duplicated and still available on another processor, i.e., $\ull u_F$ and $\ull u_{\mathcal{I}_{\Omega_{\Gamma_F}}}$ are lost but $\ull u_{\mathcal{I}_{\Omega_I}}$ and $\ull u_{\mathcal{I}_{\Omega_{\Gamma_I}}}$ are still available. Thus, it can be easily recovered by inherently using the distributed data structures. 
For the volume container, the inner node values, i.e., $\ull u_{\mathcal{I}_{\Omega_F}}$, are not available. By negelcting the interface couplings (line 3) and fixing the interface values $\ull u_{\mathcal{I}_{\Omega_{\Gamma}}}$, we numerically re-construct the lost values in $\Omega_F$ by approximating a Dirichlet problem corresponding of row 5. Asynchronously, we approximate  also a Dirichlet problem in the healthy domain corresponding to row 1. 

To coordinate the re-coupling between faulty and healthy domain, we porpose a stopping criterion for the approximation accuracy of the faulty subproblem. Therefore, we define and fix, before starting the asynchronous computation in the healthy and faulty subdomain, a bound $\sigma$ which must be fulfilled before re-coupling.  During the recovery by V-cycles in the faulty domain, we check after each iteration $\tilde k=0,1,\dots$ in the faulty domain if the local subproblem is sufficiently approximated and below this bound. For this, we use an error indicator $\eta_{\Omega_F}^{\tilde k}$. When this indicator is below the prescribed tolerance $\kappa\,\sigma$, i.e.,
\begin{equation}\label{eq:stoppingcriterion}
\eta_{\Omega_F}^{\tilde k} < \kappa\,\sigma,
\end{equation}
a signal is sent to the healthy domain to stop its computations.  In addition, we include in the bound a safety factor  $\kappa >0$. The smaller $\kappa$ is, the more difficult condition \eqref{eq:stoppingcriterion} is to satisfy.
 Asynchronously, i.e., independent of the V-cycle iteration and error estimation in the faulty domain,
 V-cycles are also executed  in the healthy domain until a signal is received from the faulty domain that the recovery has been completed.
 For this re-coupling process, the natural parallel synchronization of the multigrid  algorithm must be respected. The recovery process can signal its termination to the processes operating in the healthy domain. If such a signal is received, the healthy processes will proceed until the next canonical synchronization point of a parallel multigrid algorithm and then terminate the iteration in the respective V-cycle. More precisely, the current correction in the healthy domain is prolongated to the top
level as quickly as possible; then, the re-coupling is performed, and the regular iteration is resumed. The explict choices of the error estimator $\eta_{\Omega_F}$ and the re-coupling bound $\sigma$ will be the focus of the following sections.
 
The algorithm is presented as Algorithm \ref{alg:adaptiveDDAlg}. 
\begin{algorithm*}[h]
\caption{Multigrid cycle method with adaptively re-coupled Dirichlet-Dirichlet recovery algorithm}\label{alg:adaptiveDDAlg}
\begin{algorithmic}[1]
    \State Solve \eqref{eq:modelLS} by multigrid cycles until \eqref{def:stoppingcriterionGlobal} is satisfied.
    \If{Fault has occurred}
    \State \textbf{STOP} solving \eqref{eq:modelLS}.
    \State Recover boundary data $\ull u_{\mathcal{I}_{\Omega_{\Gamma_{F}}}}$ from line 4 in  \eqref{eq:algSystemGhost}.
    \State Initialize $\ull u_{\mathcal{I}_{\Omega_F}} $ with zero.
    \State Compute the stopping criterion for the faulty subproblem $\sigma$ and fix $\kappa>0$.
    \State \textbf{In parallel do:}
    \State ~a) \textbf{In faulty domain:} Approximate line 5 in \eqref{eq:algSystemGhost} with superman $\eta_{\super}$ by MG cycles
    \State  $\qquad A_{FF} \ull u_{\mathcal{I}_{\Omega_F}} = \ull f_{\mathcal{I}_{\Omega_F}} - A_{F\Gamma_F} \ull u_{\mathcal{I}_{\Omega_{\Gamma_F}}}$.
    \State \qquad After each cycle:
    \State \qquad Calculate the approximation indicator $\eta_{\Omega_F}$.
    \State \qquad If the stopping criterion \eqref{eq:stoppingcriterion} is fulfilled: Send signal to healthy domain and GOTO Line 21.
    \State ~b) \textbf{In healthy domain:}
    \State \qquad\textbf{In parallel do:}
    \State  \qquad\qquad i) Approximate line 1 in  \eqref{eq:algSystemGhost} by MG cycles
    \State  $\qquad\qquad\qquad A_{II} \ull u_{\mathcal{I}_{\Omega_I}} = \ull f_{\mathcal{I}_{\Omega_I}} - A_{I\Gamma_I} \ull u_{\mathcal{I}_{\Omega_{\Gamma_I}}}$.
    \State  \qquad\qquad ii) If signal was sent by faulty domain, then interrupt MG-cycle and GOTO Line 19.
   \State \qquad \textbf{end do} 
    \State  \qquad  Prolongate correction and GOTO Line 21.
    \State \textbf{end do}
    \State \textbf{RETURN} to line 1 with new values $\ull u_{\mathcal{I}_{\Omega_I}}$ in $\Omega_I$ and $\ull u_{\mathcal{I}_{\Omega_F}}$ in $\Omega_{F}$.
    \EndIf
  \end{algorithmic}
\end{algorithm*}
%
%
The adaptive controlled algorithm requires in line 1 a collective communication to inform all processes about the fault, in line 4 and 21 point-to-point communication to recover the interface data and synchronize the interface after the recovery. Additionally, we require two collective communications to determine the re-coupling criterion threshold and one for informing all processes about the re-coupling. 

The criteria \eqref{def:stoppingcriterionGlobal} for stopping the global solution approximation (see line 1 in Alg. \ref{alg:adaptiveDDAlg}) and \eqref{eq:stoppingcriterion} for the re-coupling the subproblems in the recovery (see line 12 in Alg. \ref{alg:adaptiveDDAlg})

can be realized by different approaches. 
Our explicit choices are described in Section \ref{sec:explicitcriterion}.

\subsection{Selection of the re-coupling criterion\label{sec:estimator}}
In the following subsections, we summarize the idea of the hierarchical weighted (HW) error estimator and demonstrate its efficiency for controlling the algebraic error of the iteration process, and in addition, its suitability for the recovery process by a sequence of numerical experiments. The error estimator will serve to define indicators for the error in the faulty and healthy domain.

\subsubsection{Hierarchical weighted error estimator\label{sec:HWestimator}}
%
Multigrid methods are often such efficient solvers that their computational cost, see \cite{stueben1982,brandt2011}, is lower 
than conventional error estimators, i.e.,\ a multigrid method can
compute a sufficiently accurate numerical approximation more cheaply 
than some of the error estimators that have been proposed.
Such expensive error estimators that additionally require an exact discrete solution before they can be applied, are of theoretical
interest, cf. the textbook of \cite{ainsworth2000} for a discussion on different types of estimators, but they are often unsuitable for resource-aware large scale computing.

As introduced in Section \ref{sec:MGsolver}, the $L^2$-norm of the residual of the linear system \eqref{eq:modelLS},
on level $L$
is often taken as indicator of how accurate the approximation
within an iterative scheme is, see \eqref{def:stoppingcriterionGlobal}. 
However, evaluating the residual on just one level only yields
insufficient information on the quality of an 
iterate, since 
\begin{equation}\label{eq:deterioration}
\begin{aligned}
\frac{1}{\lambda_{\max} (A_L)}\|\ull r_L^k \|_{0;L}&\leq  \|\ull e_L^k \|_{0;L}\\
& \leq \frac{1}{\lambda_{\min}(A_L)} \|\ull r_L^k \|_{0;L}
\end{aligned}
\end{equation}
with minimal eigenvalue $\lambda_{\min}(A_L) = \mathcal{O}(h_L^3)$ and maximal eigenvalue $\lambda_{\max}(A_L) = \mathcal{O}(h_L)$. The ratio of constants deteriorates with decreasing mesh size $h_L$.

In the context of multigrid methods, one obvious way
is to estimate the error by exploiting the
existing natural mesh hierarchy. 
In particular, HW error estimators
are available to measure both the algebraic error in the $L^2$- and in the $H^1$-norm,
cf. \cite{ruede1993_1}.
These estimators are efficient both in terms of lower and upper error bounds, and in
terms of run-time. 
In addition, the total error consisting of the 
algebraic and the discretization error components can be estimated,
opening up the possibility to algorithmically balance the error contributions
within adaptive control strategies, see \cite{ruede1993_2}. 
We focus in the following on the HW error estimation for the algebraic error in the $L^2$-norm.
Similar approaches can be found for the discretization error in the $H^1$-norm in the hierarchical error estimator in \cite{deuflhard1989} or for the discretization error of adaptive boundary elements in \cite{faermann1998}. 

HW error estimators are theoretically based on a stable splitting
of the underlying $H^1(\Omega)$ function space of the continuous problem.
Stable splittings can be presented in an abstract way using infinite dimensional Hilbert spaces; see \cite[Section 4.1]{oswald1994_2} for two and three dimensional problems, \cite[Chapter 2]{ruede1993_1} for two dimensions, \cite{oswald2001} for two and three dimensions. 
Since here it will be used for error estimation within the context of
the hierarchy of meshes $\mathcal{T}_0,\dots,\mathcal{T}_L$
with the associated finite dimensional approximation subspaces $V_0,\dots, V_L$,
we can work with the matrix and vector notation using the
nodal value vectors.

As a better alternative than using the residual \eqref{def:residual} for error estimation, 
one can thus use a sum of scaled residuals on a sequence of refined levels; see \cite{ruede1993_1}.
The residual of the $\ell$th linear system from \eqref{eq:modelLS} obtained by restriction of the 
$L$th residual reads
\begin{equation}\label{eq:residual}
\ull r_\ell^k \coloneqq I_L^{\ell} \ull r_L^k.
\end{equation}
Then, we define the HW error estimator by
%
\begin{equation}\label{eq:HWestimator}
\eta_{\text{HW};L} \coloneqq \left\| \sum\limits_{\ell=0}^L I_{\ell}^L D_\ell^{-1}  \ull r_\ell^k \right\|_{0;L},
\end{equation}
where $D_\ell= \diag(A_\ell)$ is the diagonal part of $A_\ell$ for $\ell > 0$ and $D_0 = A_0$. In comparison to the bounds for the residual in \eqref{eq:deterioration}, the estimator $\eta_{\text{HW};L}$ guarantees lower and upper estimates in the discrete $L^2$-norm for constants $0<c_1\leq c_2<\infty$ independent on the level $L$ 
\begin{equation}\label{eq:stable_splitting_L2}
c_1  \eta_{\text{HW};L} \leq \| \ull e_L^k \|_{0;L} \leq c_2 \eta_{\text{HW};L}.
\end{equation}
Note that $D_\ell^{-1}$ for $\ell=1,\dots,L$ in the HW estimator \eqref{eq:HWestimator} is used for scaling the residuals by $\mathcal{O}(h_\ell^{-1})$. This can be replaced by any scaling of the same order, but with the choice of $D_\ell$ also mesh informations are included. On the coarsest mesh level $\ell=0$, an exact solve of the coarsest grid problem $A_0\ull u_0 = \ull f_0$ is required. 

The proofs for two-dimensional problems from \cite[Section 3.3]{ruede1993_1} can be extended to three-dimensional problems by stable splitting arguments reported in \cite[Section 4.2]{oswald1994_2}. We additionally remark that the hierarchical sum in eq.~(\ref{eq:HWestimator}) coincides with the BPX preconditioner of \cite{bramble1990parallel}, which is derived without dimensionality restrictions.

\subsubsection{Numerical experiments\label{sec:HWnumerics}}

In the following numerical experiments, we study the HW error estimator \eqref{eq:HWestimator} for the problem and solver setup described in Section \ref{sec:exemplaryfault}.
We discretize the unit cube in a structured way by 3\,072 initial tetrahedrons and consider two test cases, one with 3 levels of refinement and a larger one having 6 levels of refinement. The corresponding fine grid problems consists then of $2.5\cdot10^5$  DOFs and $1.3\cdot10^8$ DOFs, respectively. 
To solve the discretized problem, we apply V(3,3)-cycles.
Further, on the coarsest grid level, we use a Jacobi-preconditioned conjugate gradient (PCG) method with 30 iteration steps. The number of coarse level iterations is sufficient to guarantee mesh-independent convergence of the multigrid V-cycles. Krylov subspace methods without scalable preconditioner are known to have a non-optimal complexity. However, they have been shown, e.g., in \cite{gmeiner2014,gmeiner-huber-john-ruede-wohlmuth_2015} to be a fast and sufficient method in comparison to scalable but expensive -- in terms of set-up cost -- algebraic multigrid methods for high-performance computations. For the evaluation of the estimator \eqref{eq:HWestimator}, we compute after each V-cycle iteration the residual, scale it on each level and calculate the norm. We realize the scaling in the estimator by using one application of a parallel Jacobi-smoothing routine on levels $\ell=1,\dots,L$. On the coarse level $\ell=0$, for which the inversion of $A_0$ is necessary, the same PCG-method as for the V-cycle is used. Note, that although the problem is small and an exact inversion would be possible, we only apply the same number of coarse level iterations as for the V-cycle. We observed in our experiments almost no influence on the algebraic error estimate, when solving the coarse grid problem to a higher accuracy. We note that the algebraic estimate can also be recorded in a similar form within the V-cycle iteration which is then computationally almost ``for free''.

\begin{figure*}[ht]\hspace{-2em}
\begin{minipage}{0.4\textwidth}
%
%
%
\definecolor{chocolate1}{rgb}{0.914,0.725,0.431}
\definecolor{chameleon1}{rgb}{0.541,0.886,0.204}
\definecolor{skyblue1}{rgb}{0.447,0.624,0.812}
\definecolor{plum1}{rgb}{0.678,0.498,0.659}
\definecolor{scarletred1}{rgb}{0.937,0.161,0.161}
%
		\begin{tikzpicture}[scale= 0.9]
		\begin{semilogyaxis}[
			xlabel={\# iterations for $V(3,3)$},
			ylabel={error},
			legend style={legend pos=north east,font=\scriptsize},
			yticklabel style = {font=\scriptsize},
			xticklabel style = {font=\scriptsize},
			ylabel style = {font=\small, xshift=-1.5ex},
			xlabel style = {font=\small},
			ymin=1e-17,
			ymax=1e3
		]
\addplot[color=red,mark=square*] coordinates {
(	0	,	568.919	/500.0469977912076	)	
(	1	,	42.4497	/500.0469977912076	)
(	2	,	3.9569	/500.0469977912076	)		
(	3	,	0.46291	/500.0469977912076	)
(	4	,	0.0633696	/500.0469977912076	)
(	5	,	0.00942169	/500.0469977912076	)
(	6	,	0.00146548	/500.0469977912076	)
(	7	,	0.000234864	/500.0469977912076	)
(	8	,	3.85E-05		/500.0469977912076)
(	9	,	6.43E-06		/500.0469977912076)
(	10	,	1.09E-06		/500.0469977912076)
(	11	,	1.88E-07		/500.0469977912076)
(	12	,	3.26E-08		/500.0469977912076)
(	13	,	5.72E-09		/500.0469977912076)
(	14	,	1.01E-09		/500.0469977912076)
(	15	,	1.79E-10		/500.0469977912076)
(	16	,	3.19E-11		/500.0469977912076)
(	17	,	5.69E-12		/500.0469977912076)
(	18	,	1.03E-12		/500.0469977912076)
(	19	,	2.08E-13		/500.0469977912076)
};

\addplot[color=blue,mark=o] coordinates {
(	0	,	568.919		/500.0469977912076)
(	1	,	33.5331		/500.0469977912076)
(	2	,	3.07434		/500.0469977912076)
(	3	,	0.353448		/500.0469977912076)
(	4	,	0.0477897		/500.0469977912076)
(	5	,	0.00703049		/500.0469977912076)
(	6	,	0.00108117		/500.0469977912076)
(	7	,	0.000171112		/500.0469977912076)
(	8	,	2.77E-05		/500.0469977912076)
(	9	,	4.57E-06		/500.0469977912076)
(	10	,	7.65E-07		/500.0469977912076)
(	11	,	1.30E-07		/500.0469977912076)
(	12	,	2.23E-08		/500.0469977912076)
(	13	,	3.87E-09		/500.0469977912076)
(	14	,	6.76E-10		/500.0469977912076)
(	15	,	1.19E-10		/500.0469977912076)
(	16	,	2.11E-11		/500.0469977912076)
(	17	,	3.74E-12		/500.0469977912076)
(	18	,	6.86E-13		/500.0469977912076)
(	19	,	1.87E-13		/500.0469977912076)
};

\addplot[color= skyblue1,mark=triangle*] coordinates {
(	0	,	2.70892	*17.709051617659608	)
(	1	,	8.33E-02	*17.709051617659608	)
(	2	,	4.69E-03	*17.709051617659608	)
(	3	,	3.84E-04	*17.709051617659608	)
(	4	,	4.33E-05	*17.709051617659608	)
(	5	,	6.00E-06	*17.709051617659608	)
(	6	,	9.21E-07	*17.709051617659608	)
(	7	,	1.50E-07	*17.709051617659608	)
(	8	,	2.51E-08	*17.709051617659608	)
(	9	,	4.32E-09	*17.709051617659608	)
(	10	,	7.57E-10	*17.709051617659608	)
(	11	,	1.34E-10	*17.709051617659608	)
(	12	,	2.40E-11	*17.709051617659608	)
(	13	,	4.33E-12	*17.709051617659608	)
(	14	,	7.84E-13	*17.709051617659608	)
(	15	,	1.43E-13	*17.709051617659608	)
(	16	,	2.62E-14	*17.709051617659608	)
(	17	,	5.49E-15	*17.709051617659608	)
(	18	,	2.84E-15	*17.709051617659608	)
(	19	,	2.69E-15	*17.709051617659608	)
};

\addplot[color= chocolate1,mark=triangle*] coordinates {
(	0	,	2.70892	*0.021334523340923246	)
(	1	,	8.33E-02	*0.021334523340923246	)
(	2	,	4.69E-03	*0.021334523340923246	)
(	3	,	3.84E-04	*0.021334523340923246	)
(	4	,	4.33E-05	*0.021334523340923246	)
(	5	,	6.00E-06	*0.021334523340923246	)
(	6	,	9.21E-07	*0.021334523340923246	)
(	7	,	1.50E-07	*0.021334523340923246	)
(	8	,	2.51E-08	*0.021334523340923246	)
(	9	,	4.32E-09	*0.021334523340923246	)
(	10	,	7.57E-10	*0.021334523340923246	)
(	11	,	1.34E-10	*0.021334523340923246	)
(	12	,	2.40E-11	*0.021334523340923246	)
(	13	,	4.33E-12	*0.021334523340923246	)
(	14	,	7.84E-13	*0.021334523340923246	)
(	15	,	1.43E-13	*0.021334523340923246	)
(	16	,	2.62E-14	*0.021334523340923246	)
(	17	,	5.49E-15	*0.021334523340923246	)
(	18	,	2.84E-15	*0.021334523340923246	)
(	19	,	2.69E-15	*0.021334523340923246	)
};

\legend{ $\|\ull e_L^k\|_{0;L}$, $n_{\text{HW};L}$, $ 1/\lambda_{\min;L}\| \ull r_L^k\|_{L}$,$ 1/\lambda_{\max;L}\| \ull r_L^k\|_{L}$ }
	\end{semilogyaxis}
	\end{tikzpicture}
%
 \end{minipage}
 \hspace*{4em}
 \begin{minipage}{0.4\textwidth}
%
%
%
%
%
\definecolor{chocolate1}{rgb}{0.914,0.725,0.431}
\definecolor{chameleon1}{rgb}{0.541,0.886,0.204}
\definecolor{skyblue1}{rgb}{0.447,0.624,0.812}
\definecolor{plum1}{rgb}{0.678,0.498,0.659}
\definecolor{scarletred1}{rgb}{0.937,0.161,0.161}
%
%
		\begin{tikzpicture}[scale= 0.9]
		\begin{semilogyaxis}[
			xlabel={\# iterations for $V(3,3)$},
			ylabel={error},
			legend style={legend pos=north east,font=\scriptsize},
			yticklabel style = {font=\scriptsize},
			xticklabel style = {font=\scriptsize},
			ylabel style = {font=\small, xshift=-1ex},
			xlabel style = {font=\small},
			ymin=1e-17,
			ymax=1e3
		]
\addplot[color=red,mark=square*] coordinates {
(	0	,	9931.72	/11551.312955677377)
(	1	,	732.204	/11551.312955677377)
(	2	,	67.4743	/11551.312955677377)
(	3	,	7.80956	/11551.312955677377)
(	4	,	1.06376	/11551.312955677377)
(	5	,	0.158595	/11551.312955677377)
(	6	,	0.024883	/11551.312955677377)
(	7	,	0.00403873	/11551.312955677377)
(	8	,	0.000672616	/11551.312955677377)
(	9	,	0.000114383	/11551.312955677377)
(	10	,	1.97922E-05	/11551.312955677377)
(	11	,	3.47485E-06	/11551.312955677377)
(	12	,	6.17491E-07	/11551.312955677377)
(	13	,	1.10817E-07	/11551.312955677377)
(	14	,	2.00329E-08	/11551.312955677377)
(	15	,	3.64352E-09	/11551.312955677377)
(	16	,	7.40528E-10	/11551.312955677377)
(	17	,	2.04071E-10	/11551.312955677377)
(	18	,	4.91048E-11	/11551.312955677377)
(	19	,	1.28255E-11	/11551.312955677377)
};

\addplot[color=blue,mark=o] coordinates {
(	0	,	8116.39	/11551.312955677377)
(	1	,	572.669	/11551.312955677377)
(	2	,	51.6483	/11551.312955677377)
(	3	,	5.91992	/11551.312955677377)
(	4	,	0.800424	/11551.312955677377)
(	5	,	0.118294	/11551.312955677377)
(	6	,	0.0183631	/11551.312955677377)
(	7	,	0.0029443	/11551.312955677377)
(	8	,	0.00048392	/11551.312955677377)
(	9	,	8.11816E-05	/11551.312955677377)
(	10	,	1.38581E-05	/11551.312955677377)
(	11	,	2.40115E-06	/11551.312955677377)
(	12	,	4.21376E-07	/11551.312955677377)
(	13	,	7.47535E-08	/11551.312955677377)
(	14	,	1.33852E-08	/11551.312955677377)
(	15	,	2.40146E-09	/11551.312955677377)
(	16	,	4.47217E-10	/11551.312955677377)
(	17	,	1.04014E-10	/11551.312955677377)
(	18	,	7.08505E-11	/11551.312955677377)
(	19	,	7.7725E-11	/11551.312955677377)
};
%
\addplot[color= skyblue1,mark=triangle*] coordinates {
(	0	,	1.79798	*400.14304085100144)
(	1	,	0.0511129	*400.14304085100144)
(	2	,	0.00235275*400.14304085100144	)
(	3	,	0.000129101*400.14304085100144	)
(	4	,	8.27758E-06*400.14304085100144	)
(	5	,	6.6665E-07*400.14304085100144	)
(	6	,	6.88685E-08*400.14304085100144	)
(	7	,	8.53969E-09*400.14304085100144	)
(	8	,	1.19662E-09*400.14304085100144	)
(	9	,	1.83559E-10*400.14304085100144	)
(	10	,	3.01199E-11*400.14304085100144	)
(	11	,	5.17938E-12*400.14304085100144	)
(	12	,	9.18654E-13*400.14304085100144	)
(	13	,	1.66375E-13*400.14304085100144	)
(	14	,	3.10526E-14*400.14304085100144	)
(	15	,	8.01038E-15*400.14304085100144	)
(	16	,	5.69173E-15*400.14304085100144	)
(	17	,	5.57779E-15*400.14304085100144	)
(	18	,	5.57434E-15*400.14304085100144	)
(	19	,	5.57238E-15*400.14304085100144	)
};

\addplot[color=chocolate1,mark=triangle] coordinates {
(	0	,	1.79798	*0.003715461186126746)
(	1	,	0.0511129	*0.003715461186126746)
(	2	,	0.00235275*0.003715461186126746	)
(	3	,	0.000129101*0.003715461186126746	)
(	4	,	8.27758E-06*0.003715461186126746	)
(	5	,	6.6665E-07*0.003715461186126746	)
(	6	,	6.88685E-08*0.003715461186126746	)
(	7	,	8.53969E-09*0.003715461186126746	)
(	8	,	1.19662E-09*0.003715461186126746	)
(	9	,	1.83559E-10*0.003715461186126746	)
(	10	,	3.01199E-11*0.003715461186126746	)
(	11	,	5.17938E-12*0.003715461186126746	)
(	12	,	9.18654E-13*0.003715461186126746	)
(	13	,	1.66375E-13*0.003715461186126746	)
(	14	,	3.10526E-14*0.003715461186126746	)
(	15	,	8.01038E-15*0.003715461186126746	)
(	16	,	5.69173E-15*0.003715461186126746	)
(	17	,	5.57779E-15*0.003715461186126746	)
(	18	,	5.57434E-15*0.003715461186126746	)
(	19	,	5.57238E-15*0.003715461186126746	)
};


\legend{ $\|\ull e_L^k\|_{0;L}$, $n_{\text{HW};L}$,  $ 1/\lambda_{\min;L}\| \ull r_L^k\|_{0;L}$,$ 1/\lambda_{\max;L}\| \ull r_L^k\|_{0;L}$}
	\end{semilogyaxis}
	\end{tikzpicture}
%
 \end{minipage}
  \caption{\label{fig:L2errorcomp}Convergence of the algebraic error, estimated algebraic error and scaled residual by applying V(3,3)- cycles on a box with 3\,072 initial tetrahedrons after 3 (left) and 6 (right) uniform refinement levels.}
\end{figure*}
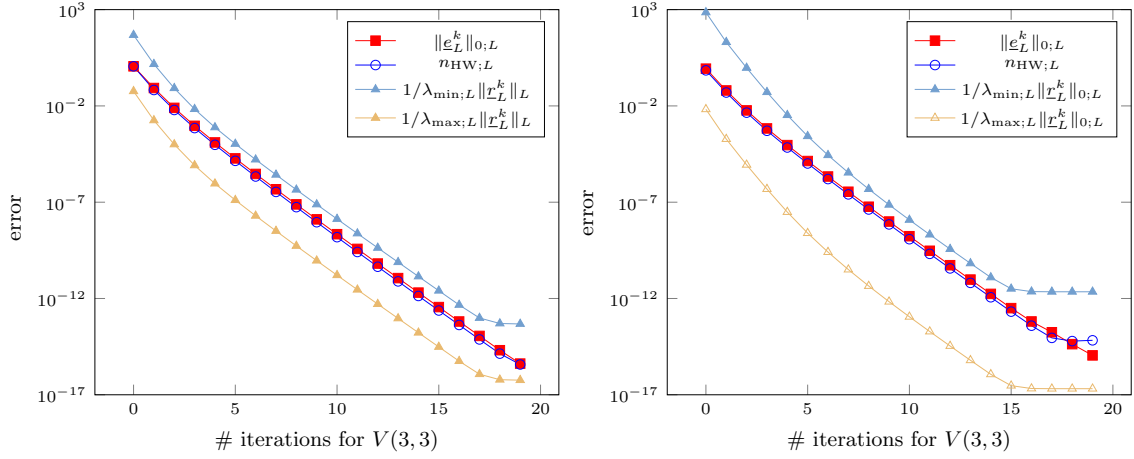

We compare the convergence of the algebraic error, the estimated algebraic error and the residual for refinement level 3 (left) and 6 (right) in Figure \ref{fig:L2errorcomp}. To compute the algebraic error, we determine in a  preprocessing step the exact finite element solution $\ull u_L$. From the estimates for the residual, see \eqref{eq:deterioration}, we scale the $L^2$-norm of the residual by the reciprocal of the minimal and maximal eigenvalue of $A_L$, which gives a lower and upper bound for the algebraic error. Therefore, we determine in another preprocessing step the extreme eigenvalues for $A_L$ for $L=3$. The computation of these eigenvalues is computationally expensive, when standard algorithms are used, and is here only feasible for moderate system sizes. These eigenvalues are then extrapolated for the large system $A_L, L=6$, by $\lambda_{\min;6} = 2^{-3}\,\lambda_{\min;3}$ and $\lambda_{\max;6} = 8^{-3}\,\lambda_{\max;3}$, where we used that the mesh is uniformly refined in each step and the knowledge about the $h_L$-scaling of the eigenvalues, see below of \eqref{eq:deterioration}.

 For both refinement levels, the residual shows in the first 3-4 iteration a pre-asymptotic convergence and then reaches its asymptotical rate. While the algebraic error convergences almost from the beginning with its asymptotic rate. The residual scaled by the extreme eigenvalue clearly bounds the algebraic error from above and below, as stated in \eqref{eq:deterioration}. However, with decreasing mesh size, the lower and upper bound deteriorate and do not provide sufficiently sharp bounds for the  algebraic error. 
When we consider the HW error estimator, we observe a very good agreement for both $L=3$, $L=6$ with the actual algebraic error in the whole convergence history.
%

\begin{figure*}[ht]
  \includegraphics[width=0.075\textwidth,trim=430mm 100mm 16mm 100mm, clip]{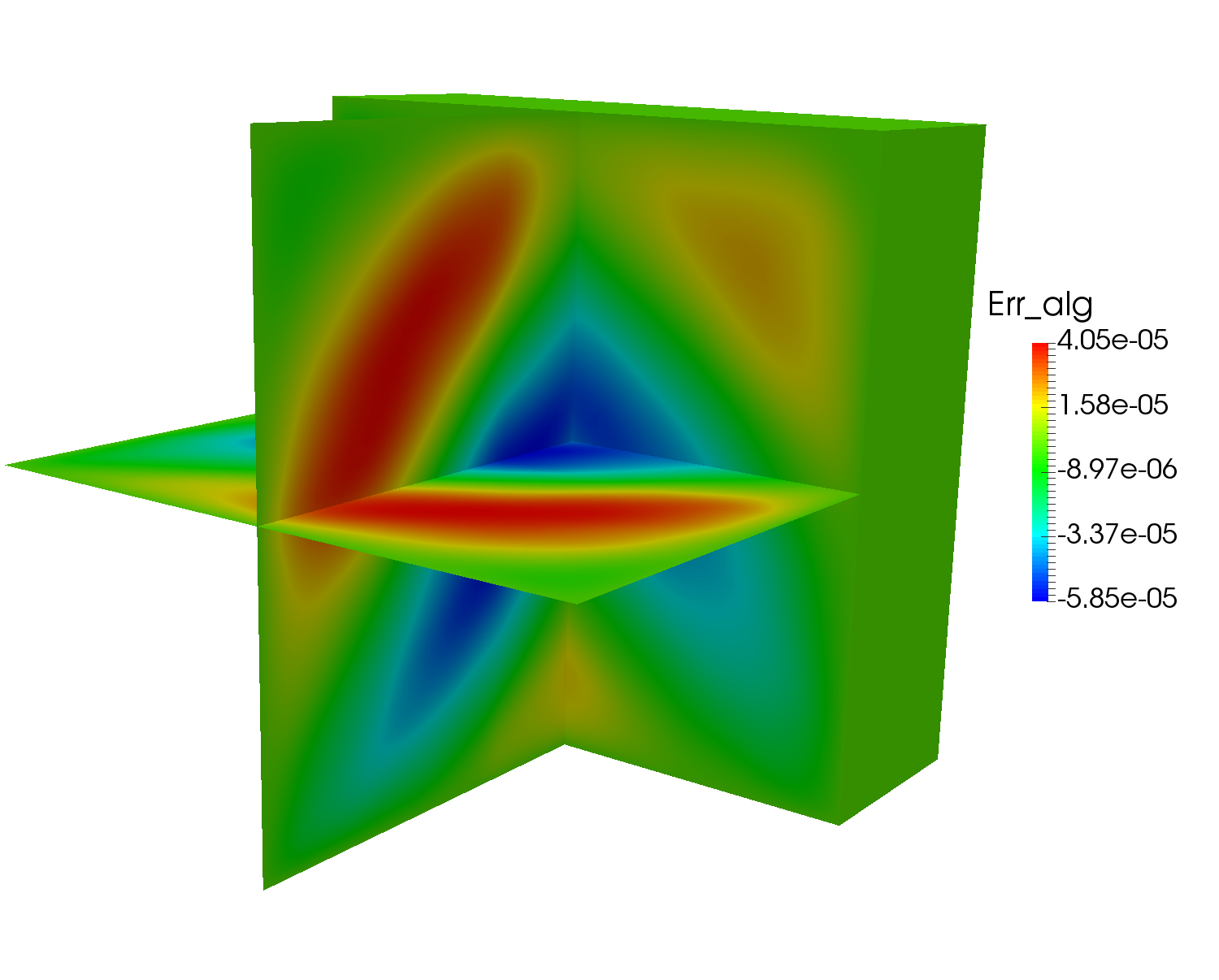}
  \includegraphics[width=0.25\textwidth,trim=10mm 5mm 105mm 5mm, clip]{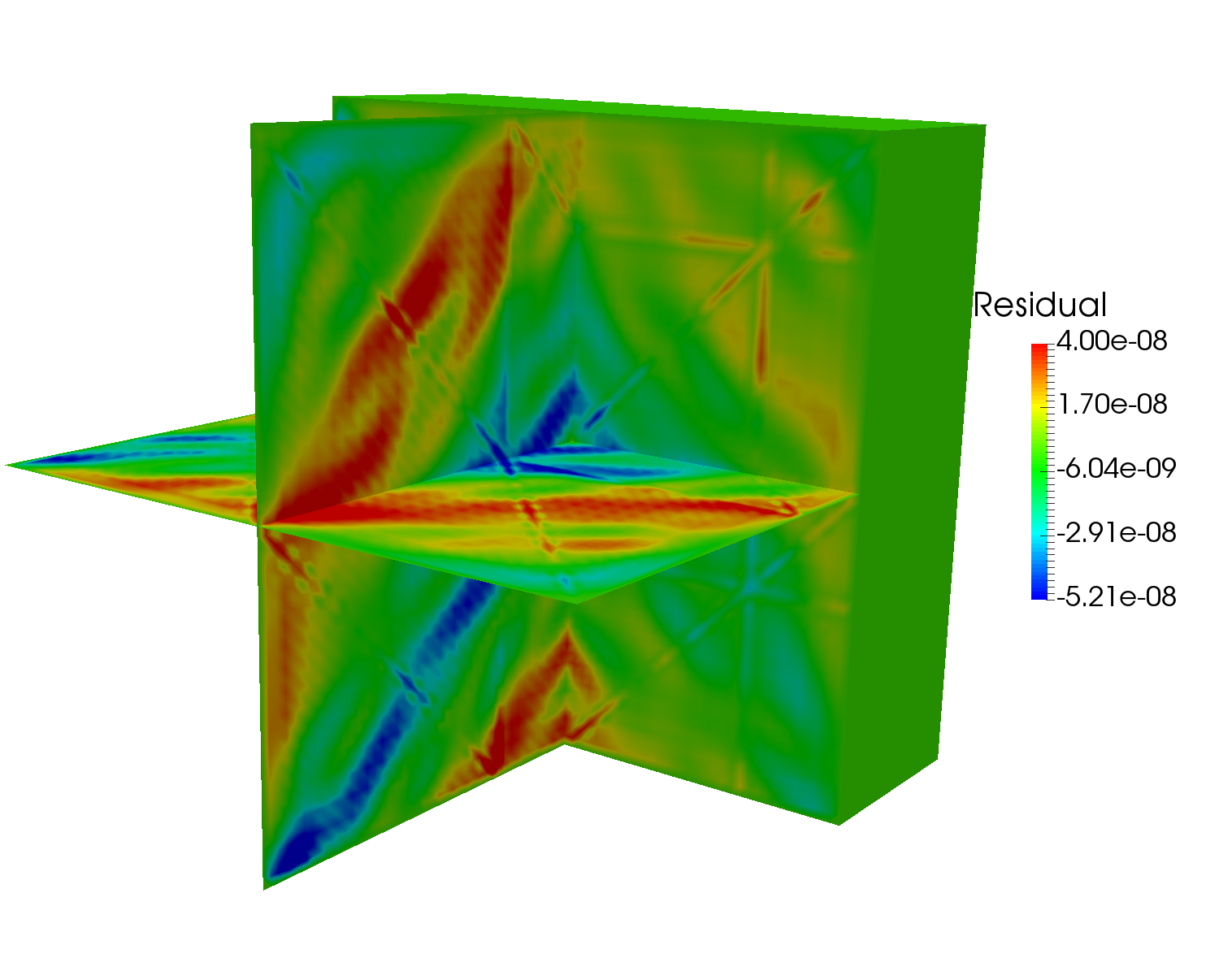}
    \includegraphics[width=0.077\textwidth,trim=424.05mm 96mm 16mm 124mm, clip]{Residual_maxLevel5_after5iter.png}

  \caption{\label{fig:L2errorlevel5}Distribution of the point-wise error components on a box with 3\,072 initial tetrahedrons and 3 uniform refinement levels after 5 V(3,3)-cycles. Left: algebraic error estimator of \eqref{eq:HWestimator} point-wise evaluated; Middle: point-wise algebraic error; Right: point-wise residual.}
\end{figure*}

In addition, we also compare the local distribution of the residual and the estimator with the algebraic error. For the recovery strategy, we are interested in the local error for which the estimator is used to define local stopping criteria. In the two series of plots in Figure~\ref{fig:L2errorlevel5} for refinement level 3, we depict the point-wise evaluation of the algebraic error in the middle after 5 V-cycle iterations.  On the right, the corresponding residual is presented   As local indicator for the $L^2$-error,  we visualize on the left the vector 
\begin{equation}\label{def:HWsum}
\sum\limits_{\ell=0}^L I_{\ell}^L D_\ell^{-1}  \ull r_\ell,
\end{equation}
which is the point-wise contribution to the HW error estimator \eqref{eq:HWestimator}.
The visualization of the algebraic error and the HW error estimator show a good visual agreement in terms of  error scales and localization. Again, the residual
is a poor indicator of the algebraic error in the size and for the error location. Furthermore, the quality of the estimates deteriorates by 3 orders of magnitude, when reducing the mesh size  from level $L=3$ to $L=6$.
%
\begin{figure*}[ht]
   \includegraphics[width=0.075\textwidth,trim=430mm 100mm 16mm 100mm, clip]{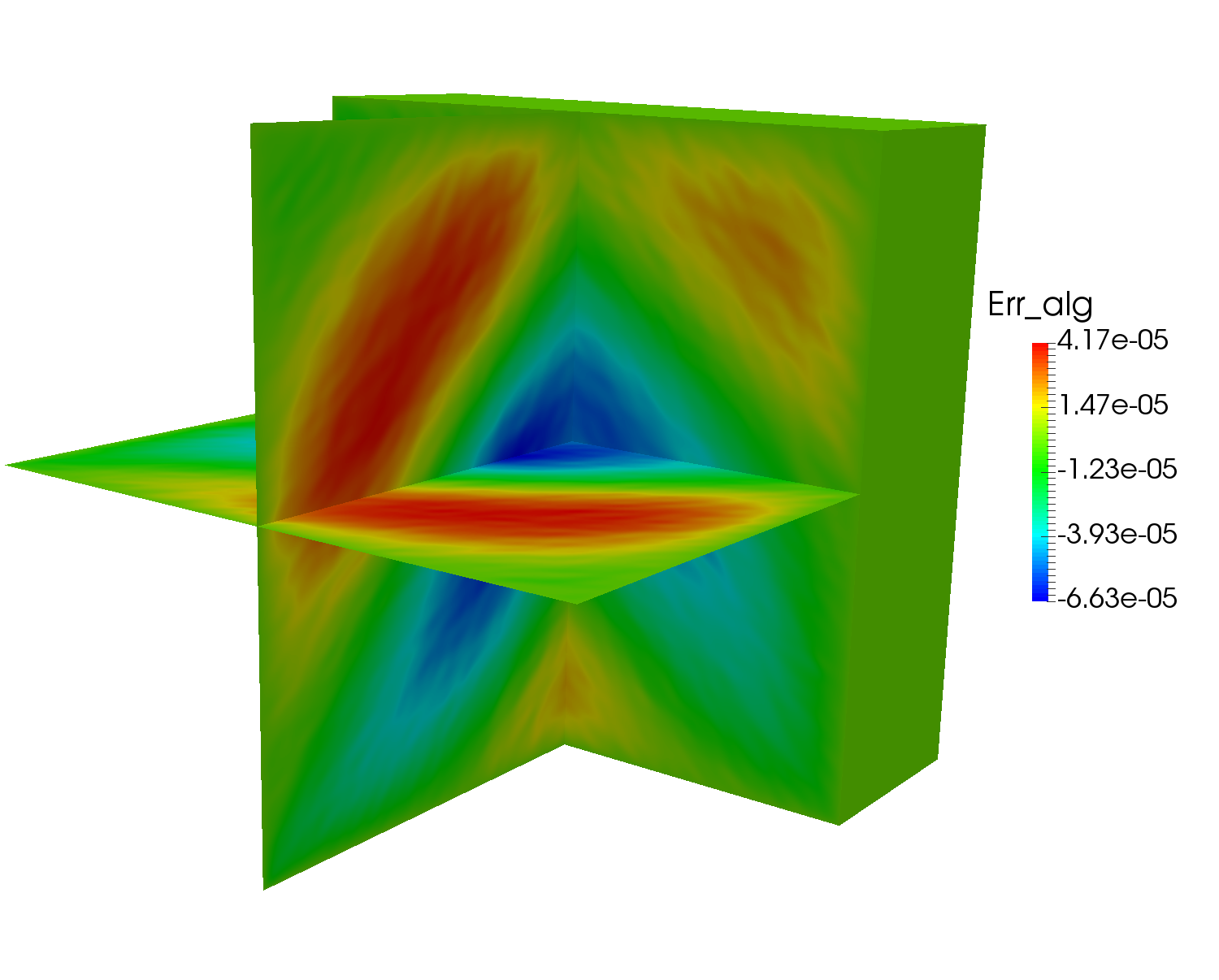}
   \includegraphics[width=0.24\textwidth,trim=45mm 5mm 115mm 5mm, clip]{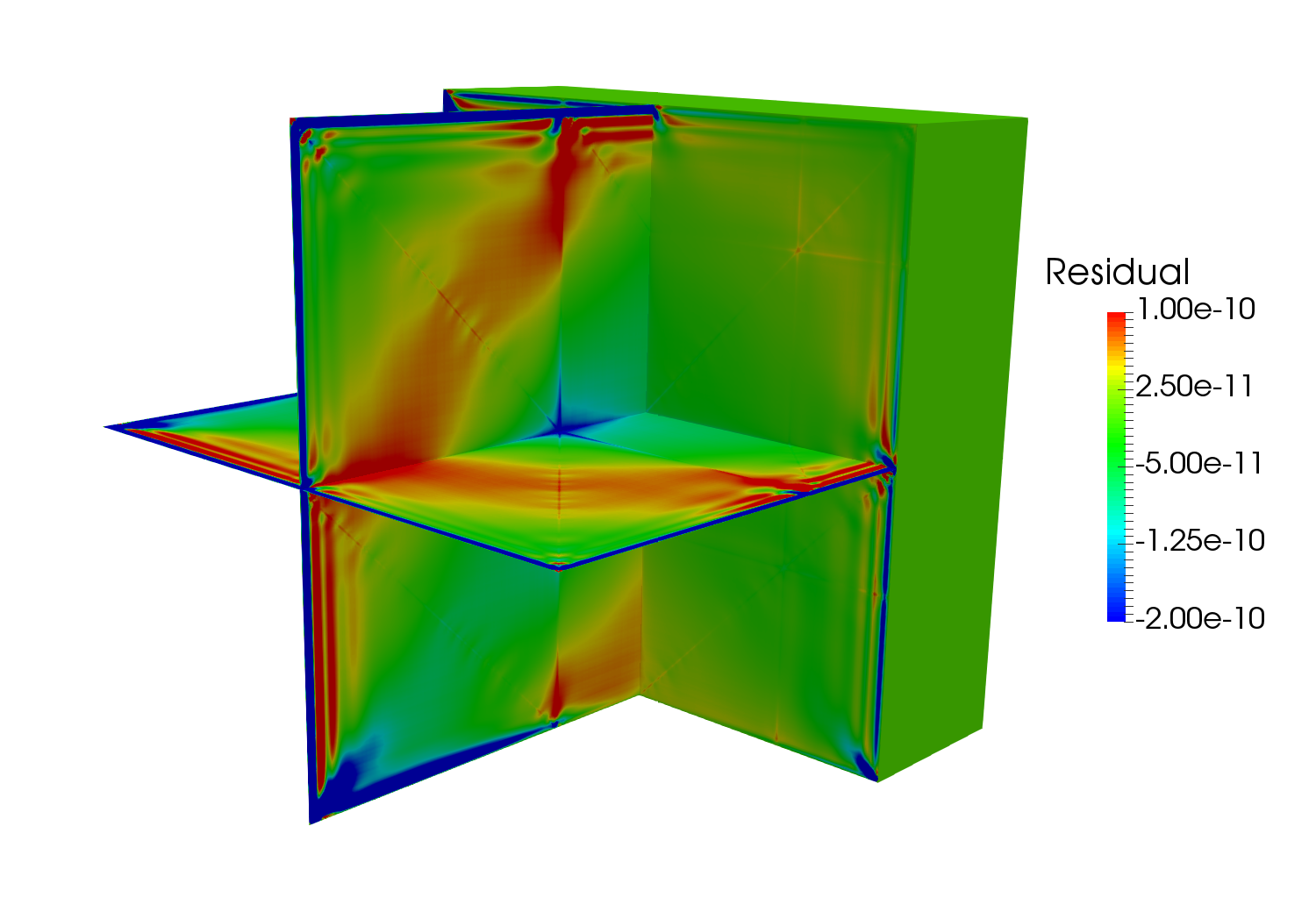}
     \includegraphics[width=0.077\textwidth,trim=420.05mm 52mm 15mm 100mm, clip]{Residual_Box_V33_5itertation_level8.png}

  \caption{\label{fig:L2errorlevel8}Distribution of the point-wise error components on a box with 3\,072 initial tetrahedrons and 6 uniform refinement levels after 5 V(3,3)-cycles. Left: algebraic error estimator of \eqref{eq:HWestimator} point-wise evaluated; Middle: point-wise algebraic error; Right: point-wise residual.}
\end{figure*}


%
\noindent


\subsubsection{Explicit choices\label{sec:explicitcriterion}}
In the following, we describe the stopping criterion choices in Alg.~\ref{alg:adaptiveDDAlg} by applying the HW estimator~\eqref{eq:HWestimator}.
In the recovery algorithm, error estimation is required at two different parts. One is necessary for the evaluation of $\eta_{\Omega}^k$ after each V-cycles iteration  in the global stopping criterion \eqref{def:stoppingcriterionGlobal} in line 1. The other one is applied for the re-coupling criterion \eqref{eq:stoppingcriterion} in line 12. In this, an efficient re-coupling bound $\sigma$ need to be defined and the computation of $\eta_{\Omega_F}$ is necessary after each  V-cycle iteration in $\Omega_F$.

In view of the previous Section \ref{sec:HWestimator}, 
the HW estimator provides better estimates than the residual. Therefore, we use in the global stopping criterion~\eqref{eq:stoppingcriterion} for estimating the algebraic error the HW~estimator 
\begin{equation}
\eta_{\Omega}^k \coloneqq \eta_{HW;L}
\end{equation}
for the approximation $\ull u^k_L$. After each V-cycle, we check if \eqref{def:stoppingcriterionGlobal} has been fulfilled to stop the iteration.

For defining suitable choices for the re-coupling bound $\sigma$ and the error indicator  $\eta_{\Omega_F}$ in the faulty subdomain, we use again the HW estimator in slightly modified form to make the computations feasible for our recovery algorithm. We re-call that each node is associated as master node to exactly one processor $p$. Let $\mathcal{I}_{\omega_p}$ be the index set of the master nodes of processor $p$ and $\mathcal{P}$ the set of processors. Then, we define the error estimator contribution $\eta_{\text{HW};\mathcal{I}_{\omega_p}}^k$ of the index set $\mathcal{I}_{\omega_p}$ to the algebraic error defined by 
\begin{equation}\label{def:ProcessContribution}
 \eta_{\text{HW};\mathcal{I}_{\omega_p}}^k \coloneqq  \left\| \sum\limits_{\ell=0}^L I_{\ell}^L D_{\ell}^{-1}  \ull r_\ell^k \right\|_{0;\mathcal{I}_{\omega_p}}.
 \end{equation}
We find a direct relation of this contribution associated with each process to the HW estimator \eqref{eq:HWestimator} by the additive decomposition 
\begin{equation}
  n_L [\eta_{\text{HW};L}^k]^2 \sum \limits_{p\in\mathcal{P}} n_{L;\mathcal{I}_{\omega_p}}[\eta_{\text{HW};\mathcal{I}_{\omega_p}}^k]^2.
\end{equation}
We note that $[\eta_{\text{HW};\mathcal{I}_{\omega_p}}^k]^2$ can be regarded as a mean quantity per node in $\mathcal{I}_{\omega_p}$, whereas $[\eta_{\text{HW};L}^k]^2$ is the mean per node for all nodes. By using \eqref{def:ProcessContribution}, we can define different  bounds $\sigma$ for the recoupling criterion \eqref{eq:stoppingcriterion}. To do so, this contribution is automatically documented and updated on each process, when estimating the algebraic error after each V-cycle application. It is additionally stored for each process and is merely a scalar quantity creating almost no overhead. 
If a failure has occured after $k=k_F$ iteration, a natural choice for the re-coupling bound $\sigma$ would be to take its contribution to the total algebraic estimate before the fault, i.e.,  
\begin{equation}
\eta_{\text{HW};\mathcal{I}_{\Omega_F}}^{k_F}.
\end{equation} 
Since all the data on the faulty domain is lost, also this value is not available, either. It could be stored by a check-pointing strategy for the estimator contribution. Here, we propose several other strategies without any necessity to introduce check-points.  
We define the follwoing two re-coupling bounds: 
The first one is based on a global mean value, and we set   
%
\begin{equation}\label{eq:stoppingcritFaultyEuc}
[\sigma^{\GRB}]^2 \coloneqq \sum_{p\in \mathcal{P}\backslash \{p_F\}} n_{L;\mathcal{I}_{\omega_p}}[\eta_{HW; \mathcal{I}_{\omega_p}}^{k_F}]^2.
\end{equation}  
We call  it {\it global mean recoupling bound} (\GRB). The second one is on the maximum contribution, and we use
\begin{equation}\label{eq:stoppingcritFaultyMax}
[\sigma^{\LRB}]^2 \coloneqq \max_{p\in \mathcal{P}\backslash \{p_F\}} n_{L;\mathcal{I}_{\omega_p}} \eta_{HW; \mathcal{I}_{\omega_p}}^2.
\end{equation}
We call it {\it local maximum recoupling bound} (\LRB).
In the setup step of the recovery, both thresholds only require a single MPI-Reduce call. 
Also other local thresholds are possible, which can be based on the minimal process-wise error or the local mean value of the process-wise error. We further point out that in the defintion of \GRB~and \LRB, the residuum $\ull r_L$ before the fault is used. A already mentioned above, this reasonable, since we store the local contribution of each process in every time the global estimate $\eta_{\text{HW};L}^k$ is computed.

Now it remains to define $\eta_{\Omega_F}$. A natural choice would be $n_{L;p_F}^{1/2} \eta_{\text{HW};\mathcal{I}_{\omega_{p_F}}}^{\tilde k}$, where now the residual of the actual recovery step $\tilde k$ is used. However, this definition is not suitable due to two reasons. Firstly, it does not necessarily tend to zero, and thus, we cannot guarantee that the stopping criterion \eqref{eq:stoppingcriterion} is ever reached. Secondly, it requires communication between different processes, and thus, it is not well suited for a low cost indicator in large scale computations. To overcome these two shortcomings, we propose to use
\begin{equation}\label{eq:HWestfaulty}
\begin{aligned}
   [\eta_{\Omega_F}^{\tilde k}]^2
  \coloneqq n_{L;\mathcal{I}_{\Omega_F}}\left \| \sum\limits_{\ell=0}^L  I_{\ell;\mathcal{I}_{\Omega_F}}^L D_{\ell; F}^{-1}  \ull r_{\ell;\mathcal{I}_{\Omega_F}}^{\tilde k} \right\|_{0;\mathcal{I}_{\Omega_F}}^2,
  \end{aligned}
\end{equation}
where $n_{L;\mathcal{I}_{\Omega_F}}$ is the number of inner nodes in $\Omega_F$, $D_{\ell;F} \coloneqq (\diag(A_{\ell}))_{FF}$ for $\ell=1,\dots,L$, $D_{0;F}\coloneqq A_{0;FF}$ and $I_{\ell;\Omega_F}^L$ is the prolongation operator from $V_{\ell;\Omega_F}$ to $V_{L;\Omega_F}$ with
\begin{equation*}
 V_{\ell;\Omega_F} \coloneqq \{  v\in H_0^1(\Omega)\colon v=w|_{\Omega_F}~ \text{for a }w\in V_\ell\},
\end{equation*}
for $\ell = 0,\dots,L$. Finally, the residual $\ull r_{\ell;\Omega_F}^{\tilde k}$ is given by $(I_{\ell;\Omega}^L)^\top \ull r_{L;\Omega_F}^{\tilde k}$, where $\ull r_{L;\Omega_F}^{\tilde k}$ is the fine scale residual restricted to all inner nodes of $\Omega_F$. It is obvious that as the number of iteration $\tilde k$ in the faulty domain tends to infinity $\ull r_{L;\Omega_F}$ tends to zero, and thus also $\eta_{\Omega_F}^{\tilde k}$. Moreover, no information exchange acorss process boundaries is required and thus no communication between proceeses. From a mathematically point of view, $\eta_{\Omega_F}^{\tilde k}$ measures the algebraic error of a local Dirichlet problem on $\Omega_F$, where the Dirichlet data are given by the values on $\Omega_{\Gamma}$ before the fault.

By these choices of re-coupling bounds and error indicator in the stopping criterion \eqref{eq:stoppingcriterion}, we can conclude the following for the error in the faulty domain after the recovery. Assuming an equally load-balanced problem, then, $n_{L;\mathcal{I}_{\omega_p}}$ is roughly the same for all processes and since $L$ is large, $n_{L;\mathcal{I}_{\omega_F}} \approx n_{L;\mathcal{I}_{\Omega_F}}$. Hence, the scaling in \eqref{eq:stoppingcritFaultyMax} and \eqref{eq:HWestfaulty} are approximately the same for all $p$. Therefore, the choice $\sigma^{\LRB}$ guarantees that after the recovery the algebraic mean error per node in the faulty domain of the surragte Dirichlet problem is bounded by the maximum mean error per process before the fault. While the bound $\sigma^{\GRB}$ guarantees that the solution of the faulty subproblem is approximated such that the mean error per node in the faulty subproblem is  of the same order  as the mean error per node  of all nodes. To further quantify the relation between the found approximation after the recovery in the faulty domain, the following bounds for \GRB~and~\LRB~are useful
\begin{equation}\label{def:boundRB}
 \sigma^{\LRB}\leq \sigma^{\GRB} \leq (|\mathcal{P}|-1)^{1/2} \sigma^{\LRB}.
\end{equation}
Hence, the difference in accuracy of the approximation quality of the faulty subproblem scales by $\mathcal{O}(|\mathcal{P}|^{1/2})$, e.g., for the largest simulations, the difference between the \GRB~and \LRB~is a factor $\mathcal{O}(10^2)$ for $29\,480$ processes and $245\,766$ processes.

We remark that a more detailed study of these stopping criterion bounds could be based on the theory of stable splittings and domain decomposition. 
By mathematically rigorous results, it may be possible to achieve the optimal re-coupling strategy, however, this may require expensive computations.
We consider the re-coupling strategy, whether it is optimal or not, in terms of its delay to reach the global stopping tolerance in case of a failure in comparison to a fault-free execution.  It will be shown that the above re-coupling bounds are suitable and of highly practical relevance.

\section{Recovery simulations in a faulty environment\label{sec:numerics}}

%
%

%
In the following section, we apply 
the adaptively controlled recovery strategy of Alg. \ref{alg:adaptiveDDAlg} with stopping criterion and error indicators of Subsec. \ref{sec:estimator}
to study the experimental fault scenario introduced in Subsec.~\ref{sec:exemplaryfault}.

Our experiments are performed on the JUQUEEN\footnote{www.fz-juelich.de} supercomputer, cf. \cite{Juqueen2015},  an IBM BlueGene/Q system with a peak performance of  $5.9$~Peta\-flop/s, listed on position 22 of the TOP500\footnote{www.top500.org} list (49th edition, Nov. 2017).  Each of the 28\,672 nodes is equipped with 16 cores that
 can execute up to four hardware supported
threads to help hiding latencies. 
The  HHG software is compiled by the IBM XL C/C++  compiler V12.1 using {\it -O3 -qstrict -qarch=qp -qtune=qp} flags and is linked to MPICH2 version 1.5.

The two stopping criteria within the recovery algorithm are specified as follows: One described by \eqref{def:stoppingcriterionGlobal} will terminate the iteration when the algebraic error of the discrete problem \eqref{eq:modelLS} estimated by the HW estimator is below $\text{TOL}=10^{-13}$. The second one described by \eqref{eq:stoppingcriterion} is used within the recovery strategy to
determine when the re-coupling should be executed so that the regular solution process can be resumed. The influence of the local stopping criterion of Section \ref{sec:explicitcriterion} will be investigated in the following numerical studies.

The fault scenario is as described in Subsec. \ref{sec:exemplaryfault}. 
The fault corrupts one MPI-process such that the data of one macro-tetrahedron ($\approx 2.7 \cdot 10^6$ DOFs) is lost, as illustrated in Fig. \ref{fig:FaultTet}.
For the algorithmic fault recovery,
we apply Alg. \ref{alg:adaptiveDDAlg}, in which the
multigrid parameters remain the same as in the global solution process, i.e.,\ V(3,3)-cycles.
Note that in the faulty domain, the coarse grid problem presents no difficulty. 
Since just one macro-tetrahedron is corrupted, 
the coarse grid consists of just a moderate number of DOFs ($\approx 35$), and thus,
this coarse grid problem can be solved efficiently with few PCG iterations.  In the healthy domain, we apply the PCG method as for the overall problem.

%
%
To realize the {\em superman} strategy, 
we partition the faulty domain further onto 2 or 4 processors
to achieve a superman speed up of approximately $\eta_{\super}=2,4$.
Thus, we assign 2 or 4  processes to perform the computations within faulty domain, 
which were handled by a single process before the fault had occurred.
The success of any recovery strategy must be measured by how much the overall solution process is delayed as compared with the fault-free case.

  
\subsection{Single Fault\label{sec:numericsexperimentsSingle}}  
\begin{figure*}[ht!]
\includegraphics[width=0.49\textwidth]{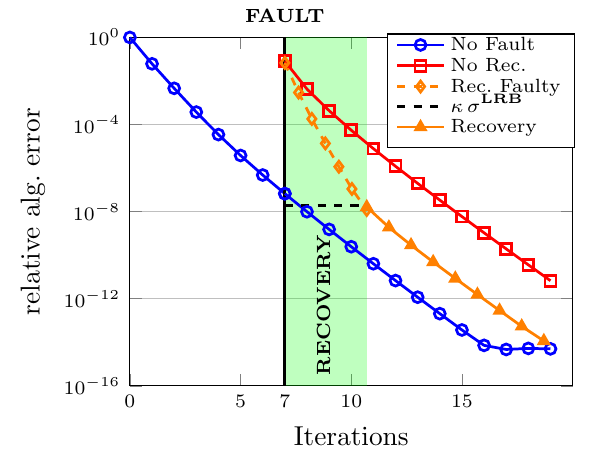}
\includegraphics[width=0.49\textwidth]{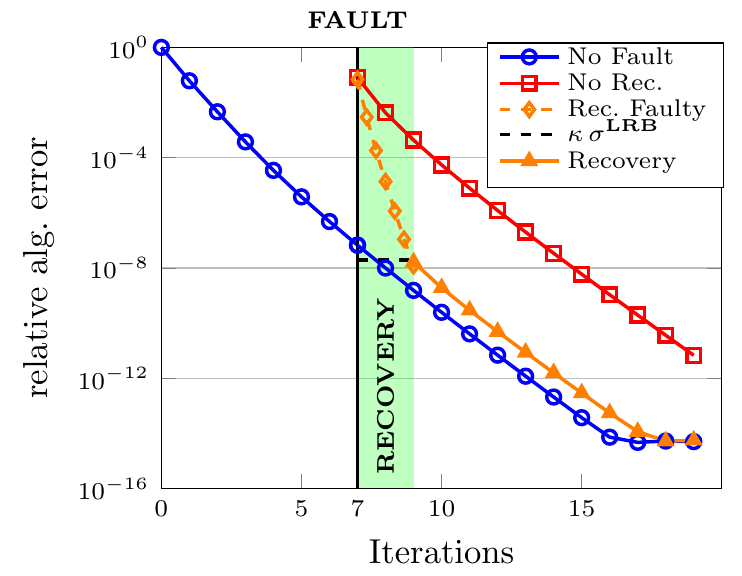}
\caption{\label{fig:adaptive_recoupling_fault_rhs}Convergence of the relative algebraic $L^2$-error obtained by the HW estimator for a fault after $k_F=7$ iterations with $\eta_{\super}= 2$ (left) and $\eta_{\super}=4$ (right) with $\sigma^{\LRB}$ and safety parameter $\kappa=1.0$.}
\end{figure*}

The input mesh in our first experiment is the unit cube discretized by 24\,576 tetrahedrons, which 
results after uniform refinement 
in a linear system for $1.3\cdot 10^{8}$ DOFs. 
Ensuring a good load-balance, 
we run these simulations with 48 MPI-processes and assign to each process 512($=8^3$) input mesh tetrahedrons. This collection of input tetrahedrons conform again a larger tetrahedron which we call macro-tetrahedron. We use in the re-coupling criterion \eqref{eq:stoppingcriterion} the bound $\sigma^{\LRB}$ and $\kappa=1.0$ for the two superman acceleration factors $\eta_{\super}=2$ and  $\eta_{\super}=4$.
In Fig.~\ref{fig:adaptive_recoupling_fault_rhs},
we present the estimated algebraic error within a faulty solution process.
The plots left and right refer to $\eta_{\super}=2$ and  $\eta_{\super}=4$, respectively.
In order to satisfy the re-coupling criterion, 6~local multigrid iterations are necessary
in the faulty subdomain which can be performed simultaneously while 3.7 and 2.0
iterations are executed in the healthy domain, respectively, depending on the superman acceleration
$\eta_{\super}$. 
In the faulty domain, the error estimator is applied in addition to the V-cycles asynchronously to the computations in the healthy domain.
The recovery process is highlighted in light green in the figures.
Note, that fractional numbers imply that the healthy domain receives the stopping signal
for re-coupling when the multigrid iteration in the healthy domain has not completed the most recent cycle, see Section \ref{sec:adaptiveStrat}.
So for example, when the healthy domain has executed the downward branch of a symmetric V-cycle, we will have executed one half of a V-cycle and denote this in the following tables and graphics by 0.5 iterations. 
Summarizing our experiments, we find that the recovery with
$\eta_{\super}=2$ results in a delay of only two to three iterations 
and $\eta_{\super}=4$ in one to two iteration delay. 
The adaptive re-coupling strategy can almost compensate for the data loss of a failing processor and can improve the global termination by up to 6 iterations.


\subsection{Influence of the local criterion and the faulty domain size\label{sec:Influencecriterion}}
In this section, we study how the algorithm behaves for the re-coupling bounds \GRB~and \LRB~including different safety parameters $\kappa$. 
To study the robustness of the re-coupling bounds for decreasing relative faulty domain sizes with respect to a fixed control parameter $\kappa$, we use a sequence of four input  meshes, consisting of $8^3\cdot6\cdot (2^m +1)^3,~m\in\{1,2,3,4\}$ tetrahedrons discretizing the unit cube.
The setup for the fault and recovery is the same as in the previous experiment, but now the relative size of the lost data decreases with each input mesh. 
The ratio between faulty and healthy domains
starts with $6\cdot 10^{-3}$ and decreases to $3.4\cdot10^{-5}$.

In the following weak scaling experiments (see Tab.~\ref{tab:recovery_performance}, \ref{tab:run_time_euc_RHS}, and \ref{tab:run_time_max_RHS}), we perform experiments in excess
of $29\,480$ parallel MPI processes.
In the scaling experiment, the number of unknowns increases asymptotically by a factor of 8 per level
so that we reach $8.2 \cdot 10^{10}$ unknowns in the largest calculation shown.
Within the multigrid scheme, we fix the mesh hierarchy size to 5 such that  the coarse grid problem grows proportionally though it is of course much smaller.
The expected growth of the coarse grid condition number is a factor of 4 for each additional
scaling level so that the number of PCG iterations is expected to grow also by $\approx \sqrt{4}$ in order to solve the coarse grid problem to the same accuracy in each level.
Thus, we compensate for this growth by doubling the number of PCG iterations
that are employed in the coarse grid solve, cf. \cite{gmeiner2014}.
For the largest simulations, we perform 240 coarse grid iterations per V-cycle.
This does not yet affect the good scalability of the multigrid method, a coarse grid solver with better scalability (such as an algebraic multigrid method)
would have to be used. 

\subsubsection{Performance of the estimator and the recovery algorithm \label{sec:perform2}}

In  Tab. \ref{tab:recovery_performance}, we present the run-times of the estimator (column {\it est}) necessary in the stopping criterion \eqref{def:stoppingcriterionGlobal} and of the asynchronously applied V-cycle in the faulty and healthy subdomain of the recovery algorithm with respect to the superman $\eta_{\super}=2$ and $\eta_{\super}=4$. Additional, we also include the run-time for error estimation (column {\it faulty} and {\it est}) necessary for evaluating \eqref{eq:stoppingcriterion}. The estimator $\eta_{\Omega}^k$ needs to be evaluated after each V-cycle iteration $k$ and $\eta_{\Omega_F}^{\tilde k}$ after each V-cycle iteration in $\Omega_F$ within the recovery.  In the estimation for the global stopping criterion and in the V-cycle iteration in healthy domain, we solve the coarse level problem by an increasing number of iterations of the coarse grid solver. Therefore, in either case the run-time increases with a parallel efficiency of 53\% from 0.86s to 1.63s for the error estimation and in the healthy domain from 2.31s to 3.17s, which constitutes a parallel efficiency of 73\%. Note in the estimator the time of the coarse solver has a much larger influence than for the run-time of the V-cycle, since on the each finer levels only one scaling is applied in estimator, while for the V-cycle 6 smoothing steps and one residual evaluation. Further, the run-time of the V-cycle in the healthy is similar to the time of one global V-cycle, since the problems have almost the same size; compare, e.g., with Tab. \ref{tab:run_time_euc_RHS}, the column fault-free and the times divided by the iterations number. We observe perfect scalability of the V-cycle and error estimation in $\Omega_F$ for both superman factors. This is rather obvious, since the size of the faulty domain remains constant in the weak scaling and has shorter communication paths, i.e., improved latency for neighboring and collective communications.  When comparing the timings for superman speed up $\eta_{\super}=2$ with $\eta_{\super}=4$, the additional compute power (by a factor of 2) is perfectly observed in the timings. Also in comparison to the run-time of V-cycle in healthy domain, run-time is reduced with respect to the speed up factor in the faulty domain. Asymptotically, the effective speed up is even large more than a factor of 3 for superman $\eta_{\super} =2$, and 6 for $\eta_{\super}=4$. This is due to the increasing run-time of the V-cycle in the healthy domain. Similarly such factors can be found for the run-times of the estimator for the global problem and for the faulty subproblem.
Summarizing, the scaling experiment shows that the HW estimator of global and faulty subproblem add less than 53\% overhead to the V-cycle application, respectively, and therefore, is suitable for  massively parallel computations within the recovery algorithm.


%
\begin{table*}[ht!]
\centering
\caption{\label{tab:recovery_performance}Timings (in sec.) of HW estimator computation (est) for the global and faulty subdomain problem. V-cycle run-times (in sec.) in the adaptive recovery for the healthy and faulty subproblem for speed up factors $\eta_{\super}=2$ and $\eta_{\super}=4$.}
\begin{tabular}{c|c||c|c|cc|cc}
\toprule
				&	&\multicolumn{1}{c|}{global}&	\multicolumn{1}{c|}{healthy}		      & \multicolumn{4}{c}{faulty}		\\
\hline
				&	&	&   &	\multicolumn{2}{c|}{Speed up $\eta_{\super}=2$} & \multicolumn{2}{c}{Speed up $\eta_{\super}=4$} \\	      
proc. &DOFs 					&  est & V-cycle &  V-cycle  & est & V-cycle & est  \\
\hline
162		&	$4.5\cdot 10^8$ 		& 0.86	  & 2.31	  & 1.01	& 0.43 	& 0.53	& 0.25		\\
750		&	$2.1 \cdot 10^9$ 		& 0.96	  & 2.44	  & 1.00	& 0.43	 & 0.51	& 0.24		\\
4\,372	&	$1.2 \cdot 10^{10}$			& 1.12	  & 2.72	  & 1.01	& 0.43	 & 0.53	& 0.25		\\
29\,480	&	$8.2 \cdot 10^{10}$			& 1.63	  & 3.17	  & 1.01	& 0.44	 & 0.54 	& 0.24			\\
\bottomrule
\end{tabular}
\end{table*}

\subsubsection{Performance of the re-coupling criterion\label{sec:performance_recoupling2}}

In the following weak scaling experiments in Tab.~\ref{tab:run_time_euc_RHS}~and~\ref{tab:run_time_max_RHS}, we study the robustness of the local re-coupling criteria for variations in $\kappa$ within the recovery, when approximating 
the solution of the linear system \eqref{eq:modelLS}. We present, here, run-times spent in the V-cycle for the global problem and for the subproblems in the recovery. We exclude the time measurements necessary for setting up the recovery in the
faulty domain and the error estimation of the global problem. The run-times for the estimation are shown in Tab. \ref{tab:recovery_performance}, in the previous Sec.~\ref{sec:perform2}. The distinction is made, since we differentiate between computation and controlling of the V-cycle iterations.

\noindent
{\it Performance of the fault-free and no-recovery simulation}\\[0.1em]
\noindent
In the fault-free execution, the iterations necessary to reach the stopping criterion stay a robust 14 within the weak scaling (in brackets in the column fault-free). The time increases slightly from 31.9s to 41.7s, which represents a parallel efficiency of more than 81\%. As for the error estimation, this observation is a consequence of using a sub-optimal coarse grid solver. In the case, when no-recovery of the data is performed after the fault and we continue with global V-cycles after re-initialization, the number of iterations necessary to satisfy the global stopping criterion  increase up to 22 iterations (in brackets in the column {\it no-recovery}).  Asymptotically (for large problems), the additionally required number of iterations decreases due to the shrinking relative size of the lost data such that the delay in convergence also decreases from 18.3s to 14.7s. 

\noindent
{\it Setup of the adaptively controlled recovery simulation}\\[0.1em]
\noindent
For the executions, in which we apply the adaptively controlled recovery in case of a fault, we report on the additional time spans which are
necessary to satisfy the global criterion.  Negative numbers indicate that the overall run-time actually improves in the case of a fault. The reason for this is that in an execution without fault the number of iterations necessary to satisfy the global stopping criterion is an integer value, while in the recovery, we allow a fractional iteration number.  In brackets, we show the number of cycles $n_F$ and $n_I$ performed in the recovery in the faulty and the healthy domain. Twice as many iterations can be executed in the faulty domain as in the healthy domain for $\eta_{\super}=2$ and
four times as many iterations for $\eta_{\super} =4$. However, we apply in the faulty subdomain in addition to the V-cycles also the error estimator. This computations are executed in the faulty domain simultaneously to the V-cycles in the healthy domain. Hence, the computation load per process is higher in the faulty domain than in the healthy one.
This is reflected in a reduced iteration number, as expected, that are performed in the faulty domain in comparison to the cycles in the healthy domain.
When increasing the problem size, the number of iterations, that can then be performed in the healthy asynchronously to the faulty one, decreases, since the run-time increases form 2.31s to 3.17s   for one V-cycle in healthy domain. Hence, for the larger problems, we observe that the factors between $n_F$ and $n_I$ increase to the expected ones.

 
 In the re-coupling criterion for the recovery, it is essential to identify a good safety parameter $\kappa$. In our extensive experiments, we have found good safety factors in a relatively wide numerical range of $\kappa \in \{10^{i},i=-2,-1,\dots,2\}$ for the re-coupling bound in \eqref{eq:stoppingcriterion}. By the choices of $\kappa$, $n_F\in[4,9]$ that is a reasonable number of iterations to recover from the fault
 for the smallest problem size. 

The re-coupling thresholds for \GRB~and \LRB~are depicted in the first column in the Tab. \ref{tab:run_time_euc_RHS} and \ref{tab:run_time_max_RHS} (bottom table), respectively. While the thresholds of \LRB~are asymptotically almost constant, \GRB~increases by a factor of 2 to 3.
By the bounds \eqref{def:boundRB}, the scaling factor between  $\sigma^{\LRB}$ and $\sigma^{\GRB}$ increases from  a factor of 13 to 172  for the largest problem, since the number of processes increases  from 162 to 29\,480 in the scaling. Consequently, when we consider a fixed $\kappa$ and increase the problem size, we observe that the number of iterations to satisfy \GRB~decreases by one iteration largest problem size, while the number of iterations for \LRB~remain constant.  Since the scaling factor between the two bounds increases rather moderately, only one iteration difference between \GRB~and \LRB is observed.

\noindent
{\it Performance with superman speed-up $\eta_{\super}=2$ and $\eta_{\super}=4$}\\[0.1em]
\noindent
Asymptotically, the same effects can be observed as in \cite{huber2016}, when decreasing the relative size between  the faulty and the healthy domain. For example, the recovery strategy can profit from the decreasing relative size and obtains more easily optimal run-times without delay in convergence. 
For both superman speed up factors, optimal run-times cannot be observed for the smallest two problem sizes and a small overhead around 2s  and 6s need to be accepted, respectively. 
The recovery is either insufficient and suggests an early re-coupling or the recovery time takes too long. In the first scenario, we observe again effects of under- and over-solving. Let us comment in more detail on the the over-solving effect. By fixing the interface data in the recovery, the error in the healthy domain can only be reduced slightly in comparison to the error in the healthy domain.
Hence, while the error in the faulty domain is still large, the error in the healthy domain already saturates. A re-coupling due to this saturation is not incorporate in the stopping criterion \eqref{eq:stoppingcriterion}, since the remaining large error components in the faulty domain can still pollute the whole domain and delay the convergence.   A higher speed up factor would be necessary to reduce this delay further. For larger problems, this effect is reduced and we can solve both subproblems to a higher a accuracy without delaying the convergence. Thus, we can slightly over-solving both problems, since the remaining large algebraic error components at the interface can be treated efficiently by the global V-cycles after the fault. Hence, the safety factor $\kappa$ can be chosen in a wider range for both re-coupling bounds and superman speed up factors. For instance for the largest problem size and the re-coupling criterion with \LRB~and $\kappa=10^{-2}$, 4 V-cycles can be executed in the healthy domain without delaying the convergence, 

The minimal run-times are observed for  \GRB~and \LRB~with $\kappa\leq 10^{-1}$ for superman speed up factor $\eta_{\super}=4$ and  for $\eta_{\super}=2$, for \GRB~and \LRB~with $\kappa= 10^{-2}$. Asymptotically, optimal run-times without delay in convergence are also obtained for \GRB~and \LRB~with $\kappa\leq 10^0$ for superman speed up factor $\eta_{\super}=4$ and for $\eta_{\super}=2$, for and \LRB~with $\kappa= 10^{-1}$.

\noindent
{\it Performance comparison of the superman speed-up factors}\\[0.2em]
\noindent
The safety factor $\kappa$ can be chosen larger for $\eta_{\super} =4$ than for $\eta_{\super} =2$, i.e., we can allow asymptotically  for the faster superman a larger error tolerance in case of re-coupling. 
The reason for this is that the superman $\eta_{\super}=4$ and the re-coupling is performed earlier than for $\eta_{\super}=2$. Therefore, the global {V-cycles} can reduce a larger error sufficiently without delay in the convergence.
For instance, consider the largest problem size and the criterion with \GRB, then, for $\eta_{\super}=4$, $\kappa$ can be chosen smaller or equal $10^0$, while for $\eta_{\super}=2$, $\kappa\leq 10^{-2}$ is necessary. This holds similar true for \LRB. Note, a too small $\kappa$ of course delays the convergence due to over-solving.

%


\noindent
{\it Performance comparison of the re-coupling criteria}\\[0.2em]
\noindent
Both re-coupling bounds perform similar well with respect to the speed up factors and can yield optimal run-times. Hence, sufficient safety parameters can obtained for smaller and larger relative size between faulty and healthy domain problems. Asymptotically, they yield robust and efficient re-coupling bound to obtain optimal run-times. 
 

When increasing the problem size and the relative ratio between faulty and healthy domain decreases, a lower approximation quality of the faulty subproblem is required. This reduced accuracy requirement is adjusted by \GRB~in the re-coupling criterion. For instance, for $\eta_{\super}=4$, an approximation of the faulty subproblem by 6 V-cycles is the minimal required iterations to obtain the best possible run-times for the smallest problem size. For the largest problem size, this are 5 V-cycles. This property can be mapped by \GRB~with $\kappa=10^0$. However, such a reduction can also lead to a too early re-coupling and need to be used carefully. For instance, while for $\eta_{\super}=2$ the safety parameter $\kappa=10^{-2}$ is identified as optimal choice, also   $\kappa=10^{0},10^{-1}$ lead the same or reasonable run-time delays. For the smallest problem size, all of these require 6 or 7 iteration in the faulty domain. Hence, it is better to choice the slightly stricter bound $\sigma^{\LRB}$.

For the largest problem size, \GRB~indicates an early re-coupling and does not leads to optimal run-times for both choices $\kappa=10^{0},10^{-1}$. It would be better to choose a smaller $\kappa$, e.g., $10^{-2}$.
When we use the \LRB~for the re-coupling, the bound does not take into account the shrinking ratio between faulty and healthy domain. Hence, asymptotically the approximation quality by the bound is higher than \GRB, see the bounds \eqref{def:boundRB}.  We can therefore say that the \LRB~slightly over-solves the subproblems in the recovery. As already state previously that asymptotically a slight over-solving of the subproblems  to a higher accuracy in the recovery does not harm the overall convergence, the \LRB~strategy leads to a more robust asymptotic behavior. For instance, when we consider the case in which the \GRB~delays the convergence, for $\eta_{\super}=2$ and $\kappa=10^{-1}$. While the delay for the \GRB~and \LRB~strategy is the same for the smallest problem, the \LRB~strategy leads to optimal run-times and the \GRB~strategy to a delayed one for the largest problem size.

\begin{table*}[ht]
\centering
\scriptsize
\caption{\label{tab:run_time_euc_RHS}Additional time spans (in sec.) of the adaptive DD recovery strategy for the re-coupling criterium with $\sigma^{\GRB}$ with
$ \kappa \in \{10^{i},i=-2,-1,\dots,2\}$
and superman speed up $\eta_{\super}=2$ (top table), $\eta_{\super}=4$ (bottom table) for a faulty
solution process $k_F=7$; time-to-solution for the fault-free execution and additional time spans; number of iterations for fault-free and no-recovery execution in brackets; number of faulty cycles $n_F$
necessary to satisfy the re-coupling criterion and corresponding healthy cycles $n_I$ in brackets for recovery simulations.
}

\begin{tabular}{c|c||c|c|ccccc}
\toprule
proc. 	&	DOFs 		& fault-free	& no recovery	&	$\kappa=10^2$ 	&	$10^{1}$ 		&	$10^{0}$ 	&	$10^{-1}$ &	$\bf 10^{-2}$ \\
\hline 
162 & $4.5\cdot  10^8$  	& 31.90   (14) & 18.30   (22) & 10.20   (4/2.4) & 9.25   (5/3.0) & 6.29   (6/3.7) & 7.50   (7/4.2) & {\bf6.56   (8/4.8)}  \\
750 &  $2.1  \cdot  10^9$ 	& 33.50   (14) & 16.80   (21) & 10.40   (4/2.3) & 9.20   (5/2.8) & 8.24   (6/3.4) & 7.06   (7/3.9) & {\bf4.31   (8/4.7)} \\
4\,372 & $1.2  \cdot  10^{10}$ 	&  36.50   (14) & 15.40   (20) & 7.99   (4/2.1) & 7.04   (5/2.7) & 3.06   (6/3.1) & 2.15   (7/3.7) & {\bf0.57   (8/4.1)} \\
29\,480 & $8.2  \cdot  10^{10}$ & 41.70   (14) & 14.70   (19) & 6.83   (3/1.4) & 5.06   (4/1.8) & 3.73   (5/2.3) & 2.38   (6/2.8) & {\bf0.73   (7/3.1)} \\ 
\bottomrule
\end{tabular}

\vspace{1em}
\begin{tabular}{c|c||c|c|ccccc}
\toprule
\multicolumn{2}{c||}{$\sigma^{\GRB}$} 					& fault-free	& no recovery	&	$\kappa=10^2$ 	&	$10^{1}$ 	&	$10^{0}$	 &	$\bf 10^{-1}$ &	$10^{-2}$ \\
\hline 
\multicolumn{2}{c||}{$2.92\cdot 10^{-4}$}	& 31.90   (14) & 18.30   (22) & 7.70   (4/1.3) & 6.26   (5/1.7) & 2.18   (6/1.9) & {\bf2.92   (7/2.2)} & 1.72   (8/2.7) \\ 
\multicolumn{2}{c||}{$2.99\cdot 10^{-4}$}	&  33.50   (14) & 16.80   (21) & 7.70   (4/1.2) & 6.50   (5/1.7) & 4.39   (6/1.8) & {\bf2.74   (7/2.1)} & 1.08   (8/2.4) \\
\multicolumn{2}{c||}{$4.29\cdot 10^{-4}$}	&   36.50   (14) & 15.40   (20) & 5.46   (4/1.1) & 3.61   (5/1.4) & -0.57   (6/1.8) & {\bf-0.13   (7/1.9)} & 0.64   (8/2.2) \\
\multicolumn{2}{c||}{$7.38\cdot 10^{-4}$}	& 41.70   (14) & 14.70   (19) & 5.16   (3/0.8) & 3.03   (4/1.0) & 0.79   (5/1.2) & {\bf-0.39   (6/1.8)} & -0.40   (7/1.8) \\ 
\bottomrule
\end{tabular}
\end{table*}

\begin{table*}[ht]
\centering
\scriptsize
\caption{\label{tab:run_time_max_RHS}Additional time spans-time (in sec.) of the adaptive DD recovery strategy for  for the re-coupling criterium with $\sigma^{\LRB}$ with
$ \kappa \in \{10^{i},i=-2,-1,\dots,2\}$ 
and superman speed up $\eta_{\super}=2$ (top table), $\eta_{\super}=4$ (bottom table) for a faulty
solution process $k_F=7$; time-to-solution for the fault-free execution and run-time delay for no-recovery and additional time span number of iterations for fault-free and no-recovery case in brackets; number of faulty cycles $n_F$
necessary to satisfy the re-coupling criterion and corresponding healthy cycles $n_I$ in brackets for recovery simulations.
}
\begin{tabular}{c|c||c|c|ccccc}
\toprule
proc. 	&	DOFs 					& fault-free	& no recovery	&	$\kappa=10^2$ 	&	$10^{1}$ 		&	$10^{0}$ 	&	$10^{-1}$&	$\bf 10^{-2}$\\
\hline
162 & $4.5\cdot  10^8$ & 31.90   (14) & 18.30   (22) & 10.20   (4/2.4) & 9.25   (5/3.0) & 6.28   (6/3.7) & 7.51   (7/4.2) & {\bf7.78   (9/5.3)} \\
750 &  $2.1  \cdot  10^9$ & 33.50   (14) & 16.80   (21)  & 9.21   (5/2.8) & 8.25   (6/3.4) & 7.06   (7/3.9) & 4.31   (8/4.7) & {\bf5.03   (9/5.0)} \\ 
4\,372 & $1.2  \cdot  10^{10}$ & 36.50   (14) & 15.40   (20)  & 8.00   (4/2.1) & 7.04   (5/2.7) & 3.04   (6/3.1) & 0.55   (8/4.1) & {\bf-0.43   (9/4.7)} \\ 
29\,480 & $8.2  \cdot  10^{10}$ & 41.70   (14) & 14.70   (19) & 5.33   (4/1.9) & 3.74   (5/2.3) & 2.37   (6/2.8) & 0.74   (7/3.1) & {\bf0.58   (9/4.0)} \\ 
\bottomrule
\end{tabular}
\vspace{1em}

\begin{tabular}{c|c||c|c|ccccc}
\toprule
\multicolumn{2}{c||}{$\sigma^{\LRB}$} 					& fault-free	& no recovery	&	$\kappa=10^2$	&	$10^{1}$ 	&	$10^{0}$ &	$\bf 10^{-1}$ &	$10^{-2}$ \\
\hline
\multicolumn{2}{c||}{$7.37\cdot 10^{-5}$}& 31.90   (14) & 18.30   (22) & 7.69   (4/1.3) & 6.26   (5/1.7) & 2.18   (6/1.9) & {\bf 2.92   (7/2.2)} & 2.20   (9/2.9) \\ 
\multicolumn{2}{c||}{$6.45\cdot 10^{-5}$}&  33.50   (14) & 16.80   (21)  & 6.50   (5/1.7) & 4.38   (6/1.8) & 2.74   (7/2.1) & {\bf 1.07   (8/2.4)} & -0.56   (9/2.7) \\ 
\multicolumn{2}{c||}{$6.04\cdot 10^{-5}$}& 36.50   (14) & 15.40   (20)  & 5.46   (4/1.1) & 3.61   (5/1.4) & -0.58   (6/1.8) & {\bf0.64   (8/2.2)} & -0.48   (9/2.7) \\ 
\multicolumn{2}{c||}{$5.82\cdot 10^{-5}$}& 41.70   (14) & 14.70   (19)  & 3.03   (4/1.0) & 0.80   (5/1.2) & -0.38   (6/1.8) & {\bf-0.41   (7/1.8)} & 0.63   (9/2.1) \\
\bottomrule
\end{tabular}
\end{table*}

\subsubsection{Local error distribution\label{sec:local_dist}}

\begin{figure*}[ht!]
 \includegraphics[width=0.5\textwidth]{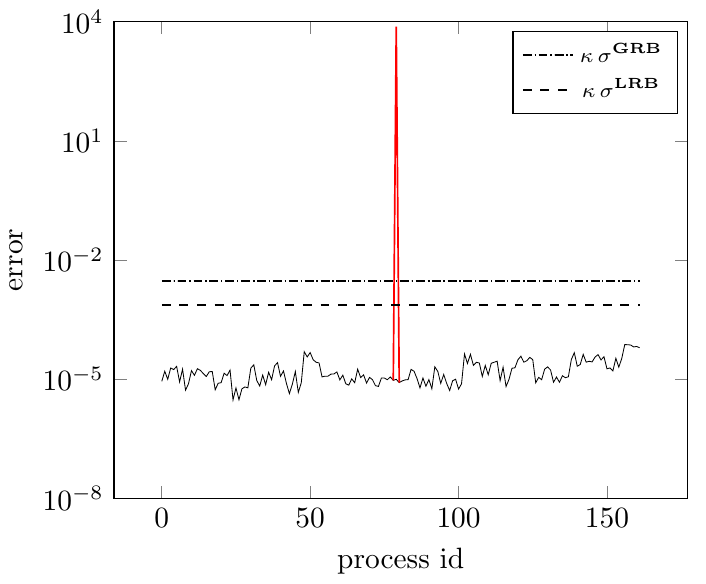}
  \includegraphics[width=0.5\textwidth]{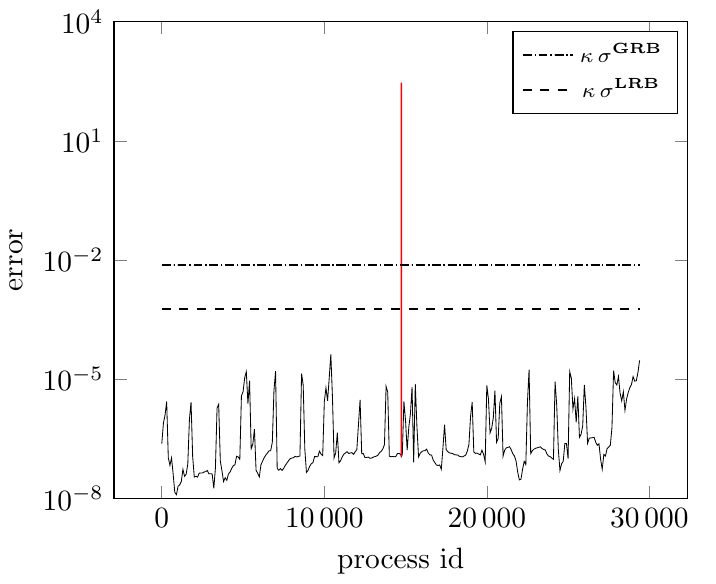}
  
   \includegraphics[width=0.5\textwidth]{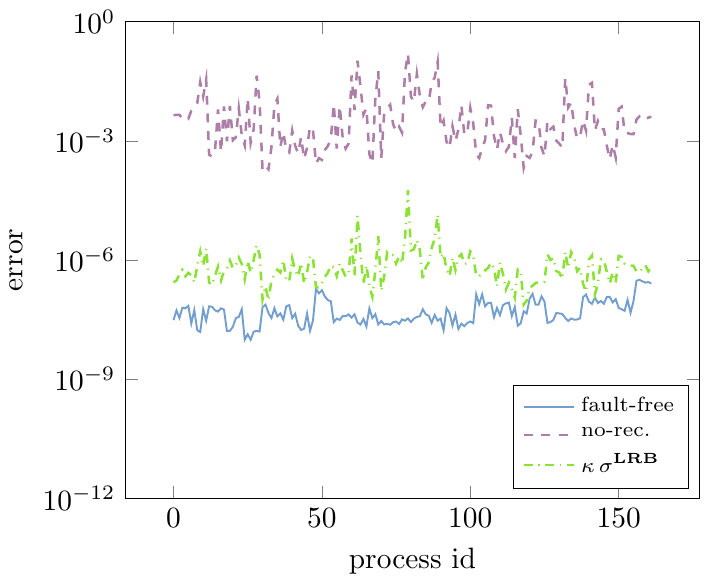}
  \includegraphics[width=0.5\textwidth]{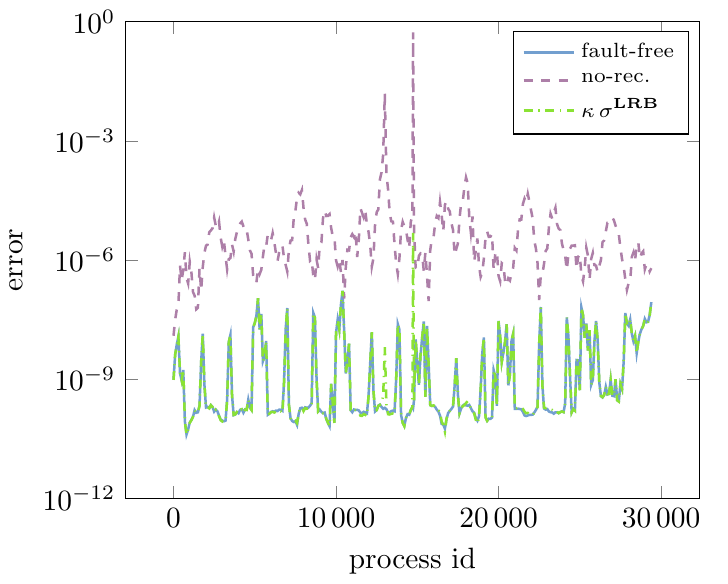}
  
     \includegraphics[width=0.5\textwidth]{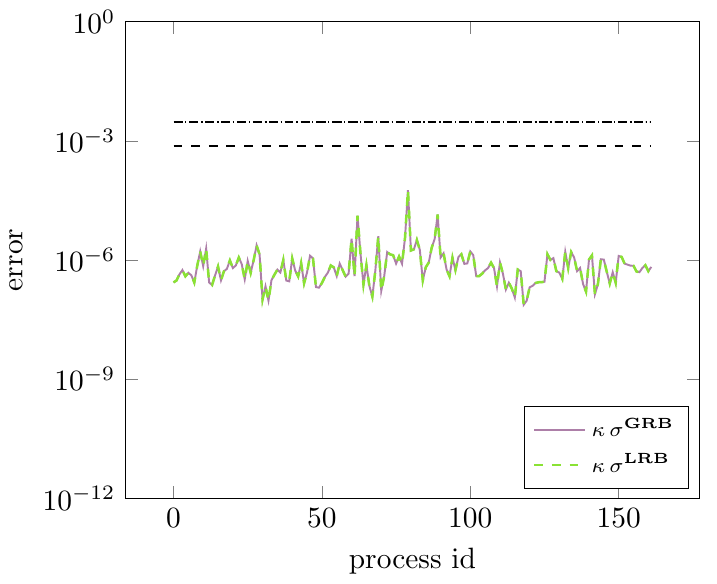}
  \includegraphics[width=0.5\textwidth]{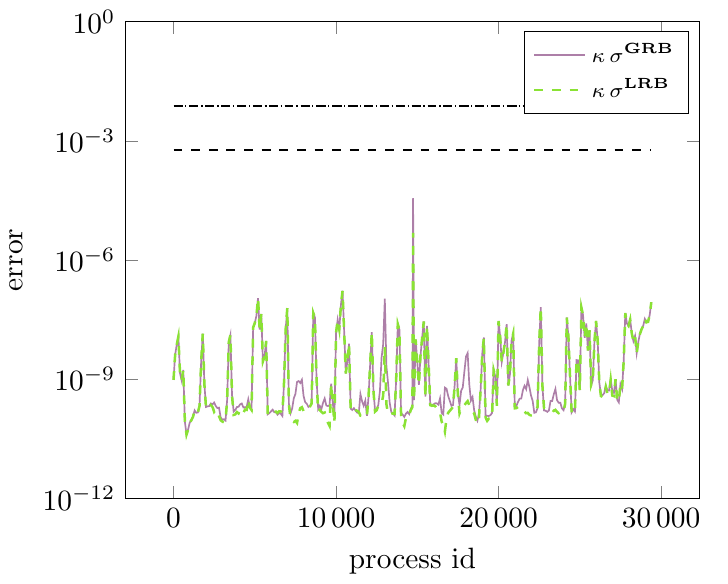}

  \caption{\label{fig:error_distribution_rhs}Process-wise distribution of the algebraic error for problem size $4.5\cdot 10^8$ DOFs on 162 processes (left) and for $8.2  \cdot  10^{10}$ DOFs on 29\,480 processes (right). Top row distribution before the fault (black), after the crash (red). Middle row, comparison of the error distribution of fault-free and no-recovery run with the adaptive controlled recovery for strategy $\sigma^{\LRB}$ with $\kappa=10^1$ and superman $\eta_{\super}=4$ after the recovery and additional applied global V-cycles. Bottom row, comparison of the process-wise error distribution of the recovery with criteria $\sigma^{\GRB}$ with $\kappa=10^1$ and for $\sigma^{\LRB}$ with $\kappa=10^1$ after the recovery and additionally applied V-cycles.}
\end{figure*}

In this section, we study the local error distribution in the case of a fault after applying a recovery strategy in more detail, and therefore, re-consider the experiments of the previous Sec. \ref{sec:performance_recoupling2}. We recall that we can relate the error contribution of each process by \eqref{def:ProcessContribution} with a portion of the global estimated algebraic error.
 We study this process-wise estimated algebraic error \eqref{def:ProcessContribution}  for the problem sizes with $4.5\cdot  10^8$ DOFs (left) and with $8.2  \cdot  10^{10}$ DOFs (right) in Fig. \ref{fig:error_distribution_rhs}.  For the largest problem, we only depict the error contribution of each 100th processor. 
 
 \noindent
{\it Error distribution before and directly after fault}\\[0.1em]
\noindent
  In the top row, the process-wise error is presented before the fault (black) and after the crash (red). 
 While the algebraic error for the smaller problem is quite equally distributed, we observe huge oscillations by more than two orders of magnitude for the larger problem. 
The reason for this is that for a zero initial guess, the difference to the exact solution is small for some DOFs, and hence, the algebraic error is small. For large process numbers, it is possible that the subdomains with exclusively small or large algebraic error components are assigned to a single process. Then, such oscillations  as seen in the illustration can occur.
By the crash, a huge local error is introduced  in the faulty domain. This error is one to two orders smaller for the larger than for the smaller problem. The DOFs per process stay constant in the scaling, but the size of the assigned subdomain is by a factor of $1/2^3$ smaller. Thus, the portion of the $L^2$-error of the process is also reduced by a factor of $(2^3)^{3/2}$. As observed in the previous Sec. \ref{sec:performance_recoupling2}, this smaller error reduces the delay in the global termination in case of a failure. We also include in these illustrations the re-coupling bounds  \GRB~and \LRB~for $\kappa=10^1$. Note, the choice of the safety parameter is not optimal, as seen by the study in the previous section. However, it serves to study the different asymptotic behavior of the bounds \GRB~and \LRB.
For the smaller problem size, both thresholds are within one order of magnitude, see the column $\sigma^{\LRB}$ and $\sigma^{\GRB}$ in Tab. \ref{tab:run_time_euc_RHS} and \ref{tab:run_time_max_RHS}. Due to this, the faulty subproblems are approximated by the same number of V-cycles.
This difference increases to more than an order of magnitude for the largest problems and the faulty subproblem is approximated by one iteration less for \GRB~than \LRB. This has for this choice of $\kappa$ also the consequence that by using \LRB~no delay is observed for the largest problem size while for \GRB~there is a delay.

\noindent
{\it Comparison of the error distribution after recovery with a fault-free and a no-recovery execution}\\[0.1em]
\noindent
In the middle row, we quantify the efficiency of the adaptively controlled recovery strategy after re-coupling for the re-coupling criterion with $\sigma^{\LRB}$ and $\kappa=10^1$ in the case of superman speed up $\eta_{\super}=4$. We consider it after the recovery in the faulty domain and one additionally global V-cycle to account for the error propagation of the remaining large error components  to the whole domain. We also include in the illustrations the error distribution in case of a fault-free and no-recovery execution. To make these runs comparable, we proceed as follows. We identify the number of performed V-cycles in the healthy domain in Tab. \ref{tab:run_time_max_RHS} in the recovery. These are 1.7 V-cycles for smaller problem and 1.2 cycles for the larger problem. We assume that one global V-cycle can be executed as fast as one V-cycle in the healthy domain such that in case of a fault-free and no-recovery execution, 
2 and 1 global V-cycles could be performed in the time of the recovery. We approximate these numbers with the closest integer value. Thus, we compare the error distribution after the recovery and one V-cycle with the error distribution of the fault-free and no-recovery execution, when 3 and 2 additional V-cycles have been performed after the fault.

In the fault-free execution, the error is equally reduced for each process by the additionally applied V-cycles. The large local error introduced by the fault pollutes the whole domain in case of the no-recovery execution such that the process-wise error is by around 5 orders of magnitude larger than in case of the fault-free execution for the smaller problem. In the larger problem, this difference is reduced to 3 orders in average. Similarly, as observed in the previous section, asymptotically, the influence of a failing process on the convergence process reduces. The largest error components in the distribution are still observed for neighboring processes of the faulty process.

The recovery strategy reduces the local large error of the faulty process to a moderate size. For the smaller problem, the process-wise error is larger than in the fault-free case. This difference is also observed in the convergence delay, see Tab.~\ref{tab:run_time_max_RHS}. For the larger problem, there exist only  small differences between the error obtained after the recovery strategy and in the fault-free case. The largest differences are observed for the faulty process by an error peak and for its neighboring process by an pollution of the remaining error after the recovery. In this case, the recovery strategy efficiently reduces all error components to the sizes comparable to the error in case of the fault-free execution such that the convergence is not delayed.

\noindent
{\it Comparison of the error distribution after recovery and global V-cycles for the re-coupling criteria}\\[0.1em]
\noindent
In the bottom row, we depict the error distribution of the adaptively controlled recovery strategy for the re-coupling criteria with $\sigma^{\GRB}$  and $\sigma^{\LRB}$ and $\kappa=10^1$. As in the previous illustration, we consider the error distribution after one additionally  V-cycle after the recovery. For the smaller problem size, $\sigma^{\GRB}$ and $\sigma^{\LRB}$ are of similar size and lead to the same number of V-cycle iteration in healthy and faulty domain in the recovery. Therefore, their error distribution after the recovery is also the same. For the larger problem, $\sigma^{\GRB}$ and $\sigma^{\LRB}$ are different and terminate their recovery with 4 and 5 V-cycles in the faulty domain and 1.0 and 1.2 V-cycles in the healthy domain, respectively. The tighter \LRB~improves the convergence such that optimal run-times are achieved, while in case of \GRB,~a delay of 3s has to be accepted, see Tab. \ref{tab:run_time_euc_RHS} and \ref{tab:run_time_max_RHS}. This is also observed in this illustration for the larger problem size. For \GRB, the process-wise error is larger than for \LRB. This is most significantly visual for the faulty and its neighboring processes. The error peak associated with the faulty process in case of \GRB~is by more than one order of magnitude larger than for the \LRB.

\subsection{Multiple Faults\label{sec:numericsexperiments}}  
\begin{figure*}[ht!]
\includegraphics[width=0.49\textwidth]{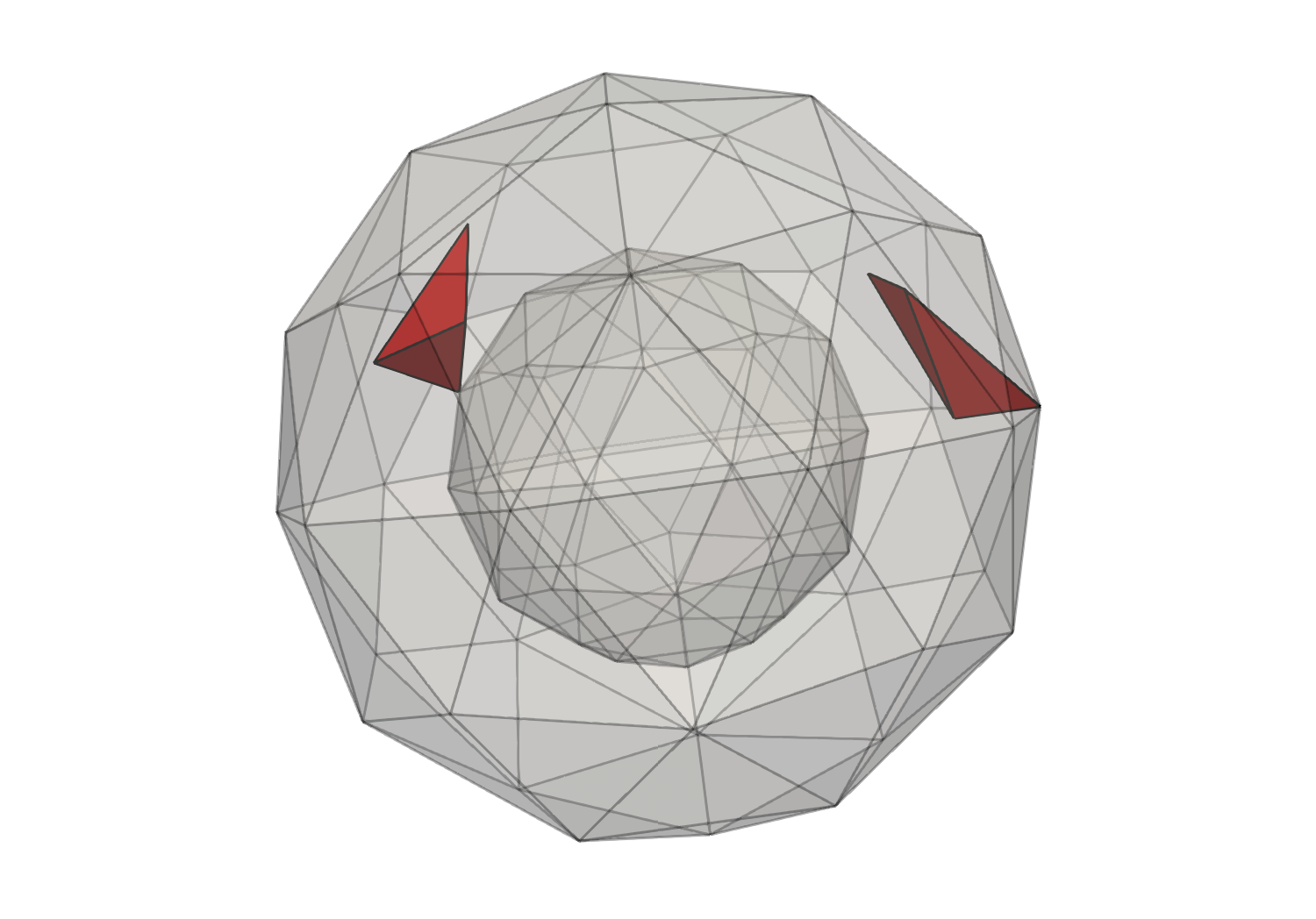}
\includegraphics[width=0.45\textwidth]{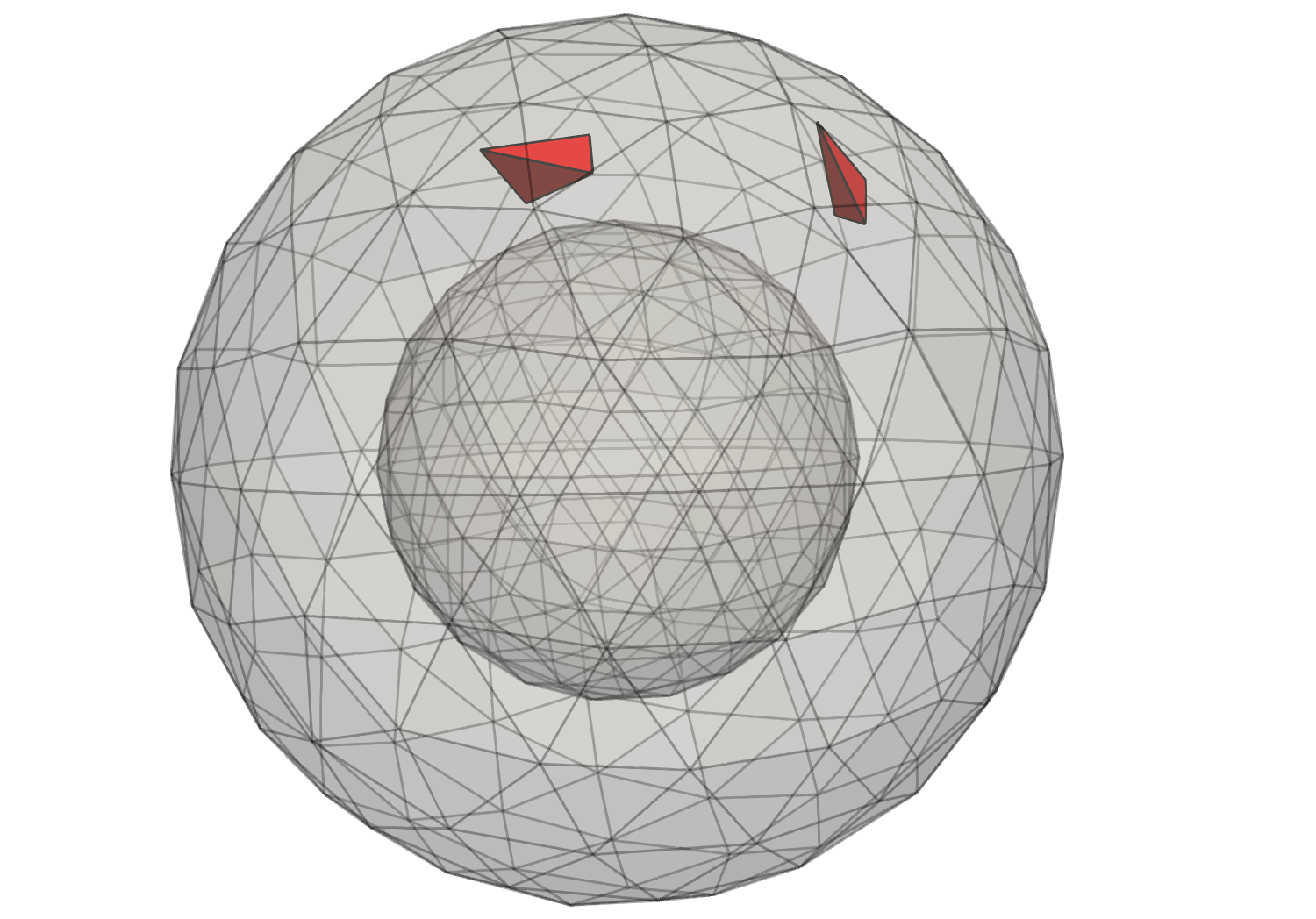}
\caption{\label{fig:sp_fault_mesh}Input mesh for different spherical scheme discretizations. The location for the two faulty processors in the initial mesh is marked by red tetrahedrons.}
\end{figure*}

In this section, we apply the adaptively controlled algorithm to a setting in which multiple faults occur in order
to demonstrate 
that the recovery algorithm is robust with respect to fault frequency variations.
For a detailed survey on models on fault probabilities and MTBF,
we refer the reader to \cite{HeraultRobert2015}.
Motivated by geophysics simulations \cite{bauer_2015}, we consider problems posed on 
a spherical shell, namely,
\begin{equation}
\Omega = \{x\in\mathbb{R}^3\colon 0.55<\|x\|<1\},
\end{equation}
where $\|\cdot\|$ denotes the Euclidean norm.
The domain $\Omega$ is discretized by an initial mesh (see Fig. \ref{fig:sp_fault_mesh}) and is then uniformly refined 5 times. Unlike in the previous experiments, 
we change the exact solution of model problem \eqref{eq:model} to 
\begin{equation}
u=\sin((x+\sqrt{2}y)\pi)\sinh(\sqrt{3}z\pi).
\end{equation}
 For this example, we have a homogenous right-hand side and prescribe the Dirichlet boundary conditions according to the exact solution.
Again, we consider the convergence of the estimated algebraic error, but 
now, we inject one fault after $k_F^1=5$ iterations and another one after $k_F^2=10$ iterations,
affecting two macro-terahedrons, as illustrated in Fig. \ref{fig:sp_fault_mesh}. 
In order to guarantee a 
load-balanced problem, we assign again to each MPI-process the same number of tetrahedron of the input mesh. In Tab. \ref{tab:run_time_multiple}, we study  a sequence of successively refined 
input meshes in order to evaluate the performance of the recovery strategy. The coarse grid grows proportionally to the fine grid while the mesh hierarchy is kept fixed. In the smallest run, the coarsest level consists of 30\,720 tetrahedrons and grows in the weak scaling to $1.3 \cdot 10^9$ tetrahedrons. 
We choose a superman acceleration of $\eta_{\super}=4$, 
since this has already shown favorable results in the context of our study of different re-coupling bounds in Sec. \ref{sec:numericsexperimentsSingle} and \ref{sec:Influencecriterion}. Moreover, the higher superman factor should also account for the late fault such that a recovery with a small delay is possible. We use  the bound $\sigma^{\LRB}$ in the re-coupling criterion and vary $\kappa \in \{10^{-1},i=-2,-1\dots,2\}$. Our largest simulation in this setup are executed on 245\,766 processes. The scaling difference between \LRB~and \GRB~is less than a factor of 500 for the largest problem size, see the bounds \eqref{def:boundRB}. The factor increase in comparison to the previous example only a factor of 3 and we expect that \GRB~performs similarly. Moreover, as indicated by the previous experiments, a slightly tighter bound as by \LRB~does not asymptotically harm the performance.
We again 
present the run-times for V-cycles 
and the additional time span spent in the no-recovery execution and spent by the recovery algorithm. 
As in the single failure experiments, the growing coarse grid and using a sub-optimal coarse grid solver require an increasing number of coarse grid iterations to compensate for the deteriorating condition number of the coarse grid stiffness matrix.
Here, we again double the number of coarse grid iterations and eventually use 320 coarse grid  PCG-iterations for the largest problem. This still constitutes only
a proportion of less than 48\% to the overall run time. 
%

%
\begin{table*}[ht!]
\centering
\small
\caption{\label{tab:run_time_multiple}Additional time span (in sec.) of the adaptive Dirichlet-Dirichlet recovery strategy for re-coupling criterion with $\sigma^{\LRB}$ and $\kappa\in \{10^{i}, i=-2,-1,\dots,1\}$, and superman speed up $\eta_{\super}=4$ for a faulty solution process $k_F^1=5$ and $k_F^2=10$;run-time for the fault-free execution and additional time for no-recovery execution; number of iterations for fault-free and no-recovery case in brackets; number of faulty cycles $n_{F_1}$ for the first recovery after $k_F^1=5$ and $n_{F_{2}}$ for the second fault after $k_F^2=10$ necessary to satisfy the stopping criterion in brackets for the recovery simulations. }
\begin{tabular}{c|c||c|c|cccc}
\toprule
proc. &	DOFs 					& fault-free	& no recovery	&	$\kappa=10^1$ 	&	$10^{0}$ 	&	$10^{-1}$	 &	$10^{-2}$ \\
\hline
60		&	$1.7\cdot 10^8$ 	& 59.0 (24)	&  20.5 (33)	&	5.9	(3,4) 	&	1.5 (5,8)	 &	2.2	(7,10)	&	2.3 (9,11)	\\
486		&	$1.3 \cdot 10^9$ 	& 40.3 (17)	&  21.4 (27)	&	7.5	(2,7)	&	5.0 (3,9)   	 &	4.3 	(4,10)	&	4.3 (5,12)	\\
3\,846	&	$1.1 \cdot 10^{10}$		& 38.3 (15)	&  22.9 (24)	&	8.1	(2,8) 	&	3.4	(3,11) 	 &	1.7   (4,11)		&	1.9	(5,13)	\\
30\,726	&	$8.5\cdot 10^{10}$		& 42.3 (15)	&  25.3 (24)	&	7.6	(2,7)	&	1.9	(3,10)  &	-0.2	(4,10)		&	2.4	(5,12)	\\
245\,766	&	$6.9\cdot 10^{11}$	& 53.0 (15)	&  27.8 (23)	&	6.5 (3,8)	& 0.5 (4,10)	&	1.1	(5,10)	&	0.6 (6,12)	\\	
\bottomrule
\end{tabular}
\end{table*}

Note that the run-times in the fault-free case are characterized by a decrease of the number of iterations for a higher mesh resolution, i.e., the  global stopping criterion \eqref{def:stoppingcriterionGlobal} with $\text{TOL}=10^{-13}$
is reached with fewer iterations on finer meshes.
This is caused by the discretization of the spherical shell geometry.  Here, a coarser initial mesh 
gives rise to less favorable element shapes that results in a worse multigrid convergence rate. A detailed study goes beyond the current article and has no direct effect on the efficiency of the fault recovery. For interested readers, we refer to the related articles, for the consideration of special smoothers in the multigrid context for unfavorable elements to \cite{gmeiner2013_1} and for the study of domains with curved boundaries in the HPC context to  \cite{bauer2017}.%

Two failures lead to an increase in the total run-time 
of up to 59\%, when no recovery is used. 
The 
automatically controlled recovery 
with safety parameters of $\kappa=10^{-1}$ and $\kappa=10^{-2}$ achieves the best results with a small overhead. 
We want to emphasize that for a late failure at, e.g., here\ $k_F^2=10$ 
the recovery must be very fast in order to achieve an optimal run-time result.
In the case, of 3\,846 processors, for example, only the time of 5 cycles 
remains until the global stopping criterion should be satisfied.
In this extreme case, a small overhead of a few seconds 
can be considered as excellent. 
The recovery  still constitutes a significant acceleration in the case of this fault situation. Note, that a larger superman acceleration factor can help to reduce this overhead further.

When the number of processes grows, the relative size of the faulty domain is reduced and
a recovery with almost no delay can be achieved with relative ease. 
For this, see, for example, the largest cases that are executed with 30\,726 or 245\,766 processes.

%
\section{Conclusion}
This paper presents a roll-forward technique that combines on-line global recovery methods within multigrid correction schemes and an adaptively steered synchronization. The on-line global recovery consists of asynchronous computations in the faulty and healthy domains. The local recovery of the faulty domain is additionally accelerated by the superman strategy. The re-coupling of the faulty and healthy subproblems is controlled by a stopping criterion for computations in the faulty domain. This stopping criterion is motivated by an error estimator that uses the underlying multigrid hierarchy structure by weighted sums of residuals and is especially suited for large scale computations. We studied the efficiency of two different stopping criteria for the re-coupling by several test cases on the BlueGene/Q peta-scale system JUQUEEN with up to $6.9\cdot 10^{11}$ DOFs on 245\,766 parallel processes. It is shown that the algorithm presented can use the automatic re-coupling  by a choice of suitable stopping criterion and recover from faults with no additional overhead in terms of run-time. 

\subsubsection*{Acknowledgements}This work was supported (in part) by the German Research Foundation (DFG) through the Priority Programme 1648 "Software for Exascale Computing" (SPP\-EXA) and through WO671/11-1.
The authors gratefully acknowledge the Gauss Centre for Supercomputing (GCS) for providing computing time through the John von Neumann Institute for Computing (NIC) on the GCS share of the supercomputer JUQUEEN at J\"ulich Supercomputing Centre (JSC).
%

\end{document}